\documentclass[usenatbib]{mn2e}

\usepackage{lscape}
\usepackage{graphicx}
\usepackage{amssymb}
\usepackage{amsmath}
\usepackage{url}
\usepackage{bm}
\usepackage{aas_macros}

\usepackage{multirow}

\title[Solar neighbourhood age scale]{A self-consistent, absolute isochronal age scale for young moving groups in the solar neighbourhood}

\author[C.~P.~M.~Bell et al.]{Cameron~P.~M.~Bell\thanks{E-mail:
  cbell@pas.rochester.edu (CPMB)},${^1}$ Eric~E.~Mamajek$^{1}$ and Tim~Naylor$^{2}$\\
$^{1}$Department of Physics \& Astronomy, University of Rochester,
Rochester, NY 14627, USA\\
$^{2}$School of Physics, University of Exeter, Exeter EX4 4QL, UK}

\begin{document}

\date{Accepted ?, Received ?; in original form ?}

\pagerange{\pageref{firstpage}--\pageref{lastpage}} \pubyear{2014}

\maketitle

\label{firstpage}

\begin{abstract}
We present a self-consistent, absolute isochronal age scale for young
$(\lesssim 200\,\rm{Myr})$, nearby ($\lesssim 100\,\rm{pc}$) moving
groups in the solar neighbourhood based on homogeneous fitting of
semi-empirical pre-main-sequence model isochrones using the $\tau^{2}$
maximum-likelihood fitting statistic of \citeauthor{Naylor06} in the
$M_V, V-J$ colour-magnitude diagram. The final adopted ages for the
groups are: $149^{+51}_{-19}\,\rm{Myr}$ for the AB Dor moving group,
$24\pm3\,\rm{Myr}$ for the $\beta$ Pic moving group (BPMG),
$45^{+11}_{-7}\,\rm{Myr}$ for the Carina association,
$42^{+6}_{-4}\,\rm{Myr}$ for the Columba association,
$11\pm3\,\rm{Myr}$ for the $\eta$ Cha cluster, $45\pm4\,\rm{Myr}$ for
the Tucana-Horologium moving group (Tuc-Hor), $10\pm3\,\rm{Myr}$ for
the TW Hya association, and $22^{+4}_{-3}\,\rm{Myr}$ for the 32 Ori
group. At this stage we are uncomfortable assigning a final,
unambiguous age to the Argus association as our membership list for
the association appears to suffer from a high level of contamination,
and therefore it remains unclear whether these stars represent a
single population of coeval stars.

Our isochronal ages for both the BPMG and Tuc-Hor are consistent with
recent lithium depletion boundary (LDB) ages, which unlike isochronal
ages, are relatively insensitive to the choice of low-mass
evolutionary models. This consistency between the isochronal and LDB
ages instills confidence that our self-consistent, absolute age scale
for young, nearby moving groups is robust, and hence we suggest that
these ages be adopted for future studies of these groups.

Software implementing the methods described in this study is available
from \url{http://www.astro.ex.ac.uk/people/timn/tau-squared/}.
\end{abstract}


\section{Introduction}
\label{introduction}

Over the past couple of decades, hundreds of young low- and
intermediate-mass stars have been discovered in close proximity to the
Sun. These stars are not uniformly dispersed across the sky, but
instead comprise sparse, (mostly) gravitationally unbound stellar
associations within which the members share a common space
motion. Approximately 10 such moving groups, with ages of between
$\simeq 10$ and 200\,Myr, have been identified within a distance of
100\,pc from the Sun (see
e.g. \citealp{Zuckerman04a,Torres08,Mamajek15}). Given their proximity
to Earth, the members of these groups therefore play a crucial role in
our understanding of the early evolution of low- and intermediate-mass
stars. Furthermore, such stars provide ideal targets for direct
imaging and other measurements of dusty debris discs, substellar
objects and, of course, extrasolar planets
(e.g. \citealp{Janson13,Chauvin15,MacGregor15}).

Whilst the relative ages of these young associations are well known
e.g. the TW Hya association (TWA) is younger than the $\beta$ Pic
moving group (BPMG), which in turn is younger than the AB Dor moving
group, the absolute ages of these groups are still
under-constrained. There are several methods for deriving the absolute
age of a given young group of stars (see the review of
\citealp{Soderblom14}), however there are known problems with almost
every one. Arguably the most common age-dating technique, using
theoretical model isochrones, still suffers from a high level of model
dependency which arises from the differences in the treatment of
various physical aspects, as well as the values of adopted parameters;
most importantly the treatment of convection, the sources of opacity,
the handling of the stellar interior/atmospheric boundary conditions
and the initial chemical composition (see e.g. \citealp{Dahm05};
\citealp*{Hillenbrand08}). Even model independent methods, such as
using kinematic information to infer the expansion rate of a given
group or estimating the time at which said group occupied the smallest
volume in space, suffer from issues of subjectivity with regard to the
stars which are included/excluded (e.g. \citealp*{Song03,delaReza06};
\citealp{Soderblom14}). Furthermore, these kinematic methods have also
been shown to be unreproducible when similar analyses have been
performed using improved astrometric data
(e.g. \citealp*{Murphy13,Mamajek14}).

Despite the apparent problems with deriving absolute ages for young
associations, recent age estimates based on the lithium depletion
boundary (LDB) technique have started to instill confidence that we
have both \emph{precise} and \emph{accurate} ages (to within just a
few Myr) for at least two of these groups -- the BPMG and
Tucana-Horologium moving group (Tuc-Hor). The LDB technique works by
identifying the lowest luminosity within a given (presumed coeval)
population of stars at which the resonant lithium (Li) feature at
$6708\,\rm{\AA}$ shows that Li remains unburned. Although the LDB
technique relies on the same theoretical models of stellar evolution
from which (model-dependent) isochrones are created, the luminosity at
which the transition from burned to unburned Li occurs is remarkably
insensitive to the inputs and assumptions noted above (see
e.g. \citealp*{Burke04,Tognelli15}).

Given this high level of model insensitivity, \cite{Soderblom14}
propose that LDB ages therefore provide the best means of establishing
a reliable and robust age scale for young ($\lesssim 200\,\rm{Myr}$)
stellar populations. Unfortunately, there is a lower age limit to the
applicability of the LDB technique ($\simeq 20\,\rm{Myr}$) which
arises due to a much higher level of model dependency between the
various evolutionary models in this age regime. Furthermore, for many
of the young groups studied here, the censuses of low-mass members is
far from complete and therefore calculating a robust LDB age is not
possible. Therefore, if we are to establish an absolute age scale for
young moving groups in the solar neighbourhood, we must use different
age-dating techniques, however we must also ensure that the resultant
ages are consistent with those calculated using the LDB technique.

In a series of papers \citep{Bell12,Bell13,Bell14} we discussed some
of the main issues with using pre-main-sequence (pre-MS) model
isochrones to derive ages for young clusters in colour-magnitude
diagrams (CMDs), and in particular, the inability of these models to
reproduce the observed loci in CMD space for clusters with
well-constrained ages and distances. In these papers we introduced a
method of creating semi-empirical pre-MS model isochrones using the
observed colours of stars in the Pleiades to derive empirical
corrections to the theoretical colour-effective temperature
($T_{\rm{eff}}$) relations and bolometric corrections (BCs; hereafter
referred to together as BC-$T_{\rm{eff}}$ relations) predicted by
atmospheric models. Of the clusters studied in the \citeauthor{Bell13}
studies, \cite{Jeffries13} recently identified the LDB in NGC~1960 and
derived an age of $22\pm4\,\rm{Myr}$ (cf. with the isochronal age of
$20\pm1\,\rm{Myr}$). The consistency between the isochronal and LDB
ages lends confidence in the use of such semi-empirical model
isochrones to establish an absolute age scale for other young clusters
and associations.

In this study we use four different sets of semi-empirical pre-MS
model isochrones (which also include the effects of binary stars) in
conjunction with a maximum-likelihood fitting statistic to derive
absolute ages for eight young moving groups within 100\,pc of the
Sun. The ages of these groups are on the same scale as those derived
for a set of young galactic clusters in \cite{Bell13}, and thus we are
beginning to establish a robust set of ages for young benchmark
stellar populations out to distances of $\simeq 1\,\rm{kpc}$. In
Section~\ref{the_data} we discuss our sample of young groups, and
having compiled a list of members and high-probability candidate
members for each, then collect broadband photometry and assign
distances. Section~\ref{the_models} introduces the theoretical stellar
interior and atmospheric models, as well as providing a brief
description of our method for creating the semi-empirical model
isochrones. In
Section~\ref{isochronal_ages_for_young_clusters_associations} we
describe the maximum-likelihood fitting statistic and derive best-fit
ages for each of the young moving groups in our sample. We then
discuss our results in Section~\ref{discussion} and finally present
our conclusions in Section~\ref{conclusions}.

\section{The Data}
\label{the_data}

We focus on the following young ($\lesssim 200\,\rm{Myr}$), nearby
($\lesssim 100\,\rm{pc}$) moving groups: the AB Dor moving group,
Argus association, BPMG, Carina association, Columba association,
$\eta$ Cha cluster, Tuc-Hor, TWA, and 32 Ori group. Assigning members
to young moving groups is typically based on a combination of
kinematic diagnostics (e.g. proper motions, radial velocities, etc.)
in conjunction with youth indicators such as H$\alpha$ and X-ray
emission, and Li absorption. Having established a list of members and
high-probability candidate members (hereafter simply referred to
together as `members') for each group, we then compile broadband
photometric measurements and assign distances to these stars so that
we can then place them in a CMD before fitting with model
isochrones. Note that in our adopted memberships for each young group
we exclude known brown dwarfs. The reason for this is that the
low-mass cut-off of the semi-empirical model isochrones we use occurs
at $\simeq 0.1\,\rm{M_{\odot}}$ (see
Section~\ref{creating_the_model_distribution}). Table~\ref{tab:member_stars}
details our list of members for each group, along with the compiled
$VJ$ photometry, spectral types and distances.

\begin{landscape}
\begin{table*}
\caption[]{$VJ$ photometry, spectral types and distances for member stars in our sample of young moving groups. The final column indicates whether the given star has a position in CMD space which is commensurate with group membership based on the best-fitting model as calculated using the $\tau^{2}$ fitting statistic (see Section~\ref{fitting_the_cmds} for details). The full table is available as Supporting Information with the online version of the paper and includes all members of the AB Dor moving group, Argus, BPMG, Carina, Columba, $\eta$ Cha, TWA, Tuc-Hor and 32 Ori.}
\centering
\begin{tabular}{l l l c c c c c c c c c c c c}
\hline
Star             &   Group   &   Sp.T.   &   Ref.   &   $V$   &   $\sigma_{V}$      &   Ref.   &   $V-J^{a}$   &   $\sigma_{V-J}$   &   Dist.   &   $\sigma_{\mathrm{Dist.}}$     &   Ref.      &   $M_{V}$   &   $\sigma_{M_{V}}$   &   $\tau^{2}$ member\\
                    &                  &             &             &   (mag)   &   (mag)                   &              &   (mag)         &   (mag)                    &   (pc)                &   (pc)                         &   (mag)        &   (mag)                          &\\
\hline
HR~8425          &        AB   &     B7IVn       &  1   &    1.734   &   0.012    &   2   &   -0.257$^{b}$  &   0.032   &   30.969   &   0.201    &   3  &   -0.721   &   0.031   &   Y\\
HR~7348          &        AB   &     B8Vs        &   1   &    3.962   &   0.004    &   2   &   -0.174$^{b}$  &   0.030   &   55.741   &   0.684    &   3   &   0.231   &   0.053   &   Y\\
HR~838            &        AB   &     B8Vn        &  1    &    3.613   &   0.016    &   2   &   -0.166$^{b}$  &   0.034   &   50.787   &   0.490    &   3   &   0.084   &   0.045   &   Y\\
HR~8911          &        AB   &     A2Vp       &   4    &    4.936   &   0.011    &   2   &   0.034$^{c}$   &   0.021   &   47.059   &   0.642    &   3   &   1.573   &   0.060   &   Y\\
HR~9016$^{\ast}$      &        AB   &     A0Va+n   &   5    &    4.568   &   0.008    &   2   &   0.036$^{c}$   &  0.022   &   42.141   &   0.391    &   3   &   1.444   &   0.041   &   Y\\
HR~1014          &        AB   &     A3V          &   6    &    6.022   &   0.009    &   7   &   0.240         &   0.025   &   54.945   &   0.906    &   3   &   2.322   &   0.072   &   Y\\
HR~7214          &        AB   &     A4V          &   8    &    5.816   &   0.009    &   7   &   0.354         &   0.022   &   54.885   &   0.934    &   3   &   2.119   &   0.076   &   Y\\
HD~25953        &        AB   &     F5V          &   9    &    7.825   &    0.012    &   7   &   0.933        &   0.029   &   55.188   &   2.802    &   3   &   4.115   &   0.221   &   Y\\
HD~4277$^{\ast}$           &        AB   &     F8V          &   10  &   7.590   &    0.014    &   11 &   0.945   &   0.024   &   52.521   &   2.455    &   3   &   3.988   &   0.204   &   Y\\
HR~1249          &        AB   &     F7V          &   12   &  5.377   &    0.011    &   2    &  0.984$^{b}$  &   0.032   &   18.832   &   0.113    &   3   &   4.002   &   0.028   &   Y\\
\hline
\end{tabular}

\vspace{1pt}
\begin{flushleft}
\textbf{Notes:}\\
Groups: (AB) AB Dor moving group; (Ar) Argus; (BP) BPMG; (Ca) Carina; (Co) Columba; (EC) $\eta$ Cha; (TH) Tuc-Hor; (TW) TWA; (32) 32 Ori\\
$^{\ast}$ denotes double or multiple star with unresolved spectral type\\
$^{a}$$J$-band photometry is from the 2MASS PSC \citep{Cutri03,Skrutskie06}, unless stated otherwise.\\
$^{b}$$V-J$ colour estimated from the $B-V$ colour due to poor 2MASS PSC photometry (see Section~\ref{data:photometry}).\\
$^{c}$$V-J$ colour estimated from the $V-K_{\rm{s}}$ colour due to poor $J$-band 2MASS PSC photometry (see Section~\ref{data:photometry}).\\
$^{d}$Distance based on weighted average of stars within the group with trigonometric distances (see Section~\ref{data:distances}).\\
\textbf{References for spectral types, photometry and distances:}
(1) \protect\cite{Garrison94}; (2) \protect\cite{Mermilliod06}; (3) Trigonometric distance from \protect\cite{vanLeeuwen07}; (4) \protect\cite{Abt95}; (5) \protect\cite{Gray87}; (6) \protect\cite{Houk75}; (7) Tycho-2 $V_{\rm{T}}$ photometry converted to Johnson $V$ following \protect\cite{Mamajek06}; (8) \protect\cite{Cowley69}; (9) \protect\cite*{Schlieder10}; (10) \protect\cite{Moore50}; (11) \protect\cite{Hauck98}; (12) \protect\cite{Gray03}; (13) \protect\cite{Torres06}; (14) \protect\cite{Houk99}; (15) \protect\cite{Gray06}; (16) \protect\cite{Kharchenko09}; (17) \protect\cite{Abt88}; (18) \protect\cite{Montes01}; (19) \protect\cite{Houk88}; (20) \protect\cite{Opolski57}; (21) \protect\cite{Yoss61}; (22) Kinematic distance derived using the `convergent point' method following \protect\citet[see Section~\ref{data:distances}]{Mamajek05}; (23) \protect\cite{Zacharias13}; (24) Kinematic distance from \protect\cite{Malo13}; (25) \protect\cite{Stephenson86}; (26) \protect\cite{Vyssotsky56}; (27) \protect\cite{Elliott14}; (28) \protect\cite*{Riaz06}; (29) Trigonometric distance from \protect\cite{Wahhaj11}; (30) Spectral type rounded to the nearest half sub-class from \protect\cite{Shkolnik09}; (31) \protect\cite{Girard11}; (32) Kinematic distance from \protect\cite{Malo14a}; (33) \protect\cite{Daemgen07}; (34) \protect\cite*{Henry94}; (35) \protect\cite{Bowler12}; (36) Trigonometric distance from \protect\cite{Shkolnik12}; (37) \protect\cite{Zacharias05}; (38) Photometric measurement and trigonometric distance from \protect\cite{Riedel14}; (39) \protect\cite{Chauvin10}; (40) \protect\cite{Gray89a}; (41) \protect\cite{Lepine11}; (42) Trigonometric distance from \protect\cite{Riedel11}; (43) \protect\cite{Lowrance00}; (44) \protect\cite{Corbally84}; (45) \protect\cite{Gray89b}; (46) \protect\cite{Pecaut13}; (47) \protect\cite{Houk82}; (48) \protect\cite{Houk78}; (49) \protect\cite{Neuhaeuser02}; (50) \protect\cite{Nefs12}; (51) \protect\cite{Pojmanski02}; (52) \protect\cite{Zuckerman04a}; (53) \protect\cite{Weis93}; (54) Kinematic distance from \protect\cite{Binks14}; (55) \protect\cite{Schroeder07}; (56) \protect\cite{Ducourant14}; (57) \protect\cite{Gray14}; (58) \protect\cite{Harlan74}; (59) \protect\cite{Lawson01}; (60) \protect\cite{Luhman04}; (61) \protect\cite{Lawson02}; (62) \protect\cite{Lyo04}; (63) \protect\cite*{Hiltner69a}; (64) \protect\cite{Mason01}; (65) Spectral type based on a combination of individual $M_{V}$ magnitudes and $B-V$ colours (see Appendix~\ref{hd20121}); (66) De-constructed colour and magnitude calculated using the technique of \protect\citet[see Appendix~\ref{photometry_split}]{Mermilliod92}; (67) \protect\cite{Levato75}; (68) Spectral type (rounded to the nearest half sub-class) and kinematic distance (adopting a 5 per cent uncertainty) from \protect\cite{Kraus14}; (69) \protect\cite*{Hawley96}' (70) \protect\cite{Costa06}; (71) \protect\cite{Weinberger13}; (72) \protect\cite{White04}; (73) \protect\cite{Webb99}; (74) \protect\cite{Barrado06}; (75) \protect\cite*{Kastner08}; (76) \protect\cite{Looper10}; (77) \protect\cite{Schneider12b}; (78) \protect\cite{Abt77}; (79) Shvonski et al. (in preparation).\\
\end{flushleft}
\label{tab:member_stars}
\end{table*}
\end{landscape}

\subsection{Members}
\label{data:members}

Our list of member stars was assembled from literature membership
lists and includes the following studies: \cite*{Mamajek99},
\cite{Luhman04}, \cite{Lyo04}, \cite{Zuckerman04a}, \cite{Mamajek05},
\cite{Lopez-Santiago06}, \cite{Torres06}, \cite{Mamajek07},
\cite{Lepine09}, \cite*{Schlieder10}, \cite{Kiss11},
\cite{Rodriguez11}, \cite{Zuckerman11}, \cite*{Schlieder12a},
\cite*{Schlieder12b}, \cite{Shkolnik12}, \cite{Binks14},
\cite{Ducourant14}, \cite{Kraus14}, \cite{Riedel14}, and Shvonski et
al. (in preparation). In addition we include the `bona fide' and
high-probability ($\geq 90$ per cent) candidate members as defined by
\cite{Malo13}, \cite{Malo14a} and \cite{Malo14b}.

The series of papers by \citeauthor{Malo13} have identified several
hundred high-probability candidate members for several of the groups
included in this study, however many of these still require additional
measurements (e.g. Li equivalent width, radial velocity, trigonometric
parallax measurement) to unambiguously assign final membership to a
given group. Given that our aim here is to derive the `best'
representative isochronal ages for young moving groups in the solar
neighbourhood, it is therefore critical that we minimise the number of
contaminating interlopers i.e. candidate members which still have a
questionable status. Therefore, for the inclusion of
\citeauthor{Malo13} high-probability candidate members, we require
that such stars must have a measured radial velocity and/or
trigonometric parallax which are/is consistent with membership for a
given group.

Not accounting for unresolved multiples, our sample includes a total
of 89 members of the AB Dor moving group, 27 members of Argus, 97
members of the BPMG, 12 members of Carina, 50 members of Columba, 18
members of $\eta$~Cha, 189 members of Tuc-Hor, 30 members of TWA, and
14 members of 32 Ori.

\subsection{Photometry}
\label{data:photometry}

The stars in our list are spread over a large area on the sky, and
although there have been dedicated searches for low-mass members in
some of the groups studied here (see e.g. \citealp{Torres06}), there
is a significant dearth of homogeneous optical photometric coverage of
these moving groups (especially in the M dwarf regime). This is not
the case in the near-infrared (near-IR), however, where all member
stars have counterparts in the Two-Micron All-Sky Survey (2MASS;
\citealp{Skrutskie06}) Point Source Catalog (PSC; \citealp{Cutri03}).

Although ages can be derived using near-IR CMDs, the loci become
vertical with colours $J-K_{\rm{s}} \simeq 0.9\,\rm{mag}$ for $2500
\lesssim T_{\rm{eff}} \lesssim 4000\,\rm{K}$. So although stars become
less luminous as a function of time, the sequences for different
groups are almost degenerate with age. Furthermore, at very young ages
($\lesssim 10\,\rm{Myr}$), observations are further complicated by the
presence of circumstellar material. Therefore, to derive ages for our
sample of groups we supplement the near-IR photometry with $V$-band
data. Note that due to increased contamination in the
$K_{\rm{s}}$-band photometry as a result of circumstellar material
(especially for $\eta$ Cha and TWA), we derive ages using the $M_{V},
V-J$ CMD.

Whilst many of the stars in our list of members have \emph{Hipparcos}
entries, this catalogue, unfortunately, does not quote explicit
uncertainties on the $V$-band magnitude. We have therefore assembled
$V$-band photometry from the following sources: \cite{Vyssotsky56},
\cite{Hauck98}, Tycho-2 \citep{Hog00b} $V_{\rm{T}}$ transformed into
Johnson $V$ using the relation of \cite*{Mamajek06}, \cite{Lawson01},
\cite{Lawson02}, the All-Sky Automated Survey (ASAS;
\citealp{Pojmanski02}), \cite{Lyo04}, \cite{Zuckerman04a}, the Naval
Observatory Merged Astrometric Dataset (NOMAD; \citealp{Zacharias05}),
\cite{Barrado06}, \cite{Costa06}, \cite{Mermilliod06}, the Search for
Associations Containing Young stars (SACY) sample \citep{Torres06},
the All-Sky Compiled Catalogue of 2.5 million stars (ASCC-2.5;
\citealp{Kharchenko09}), \cite{Chauvin10}, the Southern Proper Motion
Catalog 4 (SPM4; \citealp{Girard11}), \cite{Lepine11}, the USNO CCD
Astrograph Catalog 4 (UCAC4; \citealp{Zacharias13}), and
\cite{Riedel14}.

Whilst the majority of these studies provide associated uncertainties
on the $V$-band magnitude, stars for which we have adopted NOMAD and
SPM4 photometry, as well as photometry taken from the studies of
\cite{Zuckerman04a}, \cite{Barrado06}, \cite{Chauvin10}, and
\cite{Lepine11} do not. In such cases we adopt a conservative $V$-band
uncertainty of $\sigma_{V}=0.1\,\rm{mag}$. Following \cite{Pecaut13},
for SACY objects brighter than $V=12\,\rm{mag}$ we adopt an
uncertainty of $\sigma_{V}=0.01\,\rm{mag}$, whereas for fainter
sources we again adopt the rather conservative
$\sigma_{V}=0.1\,\rm{mag}$.

There are two issues which we must consider before placing a given
star in the CMD, namely unresolved binary stars and low-quality 2MASS
PSC photometry. Addressing the former, our list of members naturally
includes a significant number of binaries, and although for several of
these both components are resolved in the $V$- and $J$-bands, for
others, the optical and/or near-IR photometry is unresolved. If the
two components are resolved in both the $V$- and $J$-bands, then we
plot each component separately in the CMD. However, given that we have
heterogeneous information concerning the binary population of all the
co-moving groups (e.g. the lack of information concerning the swathe
of new Tuc-Hor members from \citealp{Kraus14}) coupled with the fact
that the model isochrones we use to derive ages include the effects of
binary stars (see Section~\ref{fitting_the_cmds}), then if either the
$V$- and/or $J$-band photometry is unresolved we instead opt to plot
the combined system measurement in the CMD. Note that the model
isochrones do not account for higher order multiples (e.g. triples or
quadruples), and so if we have information concerning the number of
components for such systems (e.g. TWA~4 is a quadruple system) we then
attempt to derive individual component colours and magnitudes
following the technique of \cite{Mermilliod92} based on the system
colour and magnitude as well as the magnitude difference between the
components in a given photometric bandpass. This process is explained
fully in Appendix~\ref{photometry_split} and specifies the systems for
which this technique was adopted. In the case of higher order multiple
systems for which we do not have the required information to derive
individual component colours and magnitudes (e.g. the magnitude
difference between the various components) we simply plot the combined
system measurement in the CMD.

Secondly, to ensure we are using the best possible photometry from the
2MASS PSC, we only use those objects for which the associated Qflg is
`A'. For many of the brighter, early-type stars, however, the
associated Qflg is `C', `D' or `E' i.e. either a magnitude was
extracted but the associated uncertainty is prohibitively large or
there were serious issues extracting a magnitude at all. For several
cases, whilst the $K_{\rm{s}}$-band photometry has an associated Qflg
of `A', the $J$-band photometry does not, and so we therefore use the
main-sequence relation of \cite{Pecaut13} to infer a $V-J$ colour for
the star based on its $V-K_{\rm{s}}$ colour. For a smaller number of
cases all three 2MASS PSC bands have associated Qflgs which are not
`A', and in such instances we instead use the associated $B-V$ colour
(typically from \citealp{Mermilliod06}) to infer the $V-J$ colour
again using the \cite{Pecaut13} relation. The stars which are affected
as such are expected to have settled on the main-sequence (based on
their spectral type) and so we can be confident that such a technique
will yield colours which can be used to reliably place stars in the
CMD. A full list of the stars for which we have adopted this
boot-strapping of colours, in addition to objects with questionable
memberships and those we have excluded from our age analysis, can be
found in Appendix~\ref{notes_on_individual_stars}.

\subsection{Distances}
\label{data:distances}

We preferentially adopt trigonometric parallax measurements for
assigning distances to each star, however in cases where these are not
available or the associated uncertainties are prohibitively large, we
adopt either a kinematic distance from the literature or derive one
using the so-called `convergent point' method
\citep{deBruijne99,Mamajek05}. In certain cases, the trigonometric
parallaxes are possibly erroneous (e.g. TWA~9A and TWA~9B where the
\emph{Hipparcos} parallax may be in error at the $\sim 3\sigma$ level;
see \citealp{Pecaut13}), and therefore we adopt a kinematic distance
instead. Of the young groups in our sample we require kinematic
distances for several members of the AB Dor moving group, BPMG,
Tuc-Hor and TWA, for which we adopt the convergent point solutions of
\cite{Barenfeld13}, \cite{Mamajek14}, \cite{Kraus14} and
\cite{Weinberger13} respectively.

For the two most distant groups in our sample -- $\eta$ Cha and 32 Ori
-- there are only a small number of stars with trigonometric parallax
measurements (RECX~2 [$\eta$~Cha] and RECX~8 [RS~Cha] in $\eta$ Cha
and 32~Ori, HR~1807 and HD~35714 in 32 Ori). As opposed to deriving
individual kinematic distances for the other members in these groups,
we instead adopt distances of $94.27\pm1.18\,\rm{pc}$ and
$91.86\pm2.42\,\rm{pc}$ for stars in $\eta$ Cha and 32 Ori
respectively based on the weighted average of stars within these
clusters with trigonometric parallaxes from the revised
\emph{Hipparcos} reduction of \cite{vanLeeuwen07}.

\section{The Models}
\label{the_models}

In this study we adopt four sets of semi-empirical model isochrones
based on different interior models, namely the \cite{Dotter08},
\cite{Tognelli11}, \cite{Bressan12} and \cite{Baraffe15} models
(hereafter referred to as Dartmouth, Pisa, PARSEC and BHAC15
respectively)\footnote{The semi-empirical Dartmouth and Pisa models we
  use here are available via the Cluster Collaboration isochrone
  server
  \url{http://www.astro.ex.ac.uk/people/timn/isochrones/}}. Note that
the Pisa models represent a customised set of interior models which
span a much greater mass and age range than those available via the
Pisa webpages (see \citealp{Bell14} for details). In addition, we note
that the semi-empirical PARSEC models are \emph{not} the recent v1.2S
models by \cite{Chen14}. Instead, they are based on the v1.1 interior
models (computed assuming a solar composition of $Z=0.0152$) and were
created using an identical process to the other semi-empirical model
isochrones; a process briefly summarised below.

It has been demonstrated that using atmospheric models to transform
theoretical interior models into CMD space results in model isochrones
that do not match the shape of the observed locus, especially at
low-masses (see e.g. \citealp{Bell12,Lodieu13}). Even with the use of
empirical BC-$T_{\rm{eff}}$ relations, the aforementioned discrepancy
between the models and the data still exists, especially in optical
bandpasses (see e.g. \citealp{Bell14}). Hence, in \cite{Bell13,Bell14}
we described a method of calculating additional empirical corrections
to the theoretical BC-$T_{\rm{eff}}$ relations predicted by the
BT-Settl atmospheric models \citep{Allard11} so as to match the
observed colours of low-mass stars in the Pleiades assuming an age of
$130\,\rm{Myr}$, distance modulus of $5.63\,\rm{mag}$ and reddening of
$E(B-V)=0.04\,\rm{mag}$. These corrections were then combined with the
theoretical dependence of the BCs on surface gravity from the
atmospheric models to derive semi-empirical BC-$T_{\rm{eff}}$
relations.

\begin{figure}
\centering
\includegraphics[width=\columnwidth]{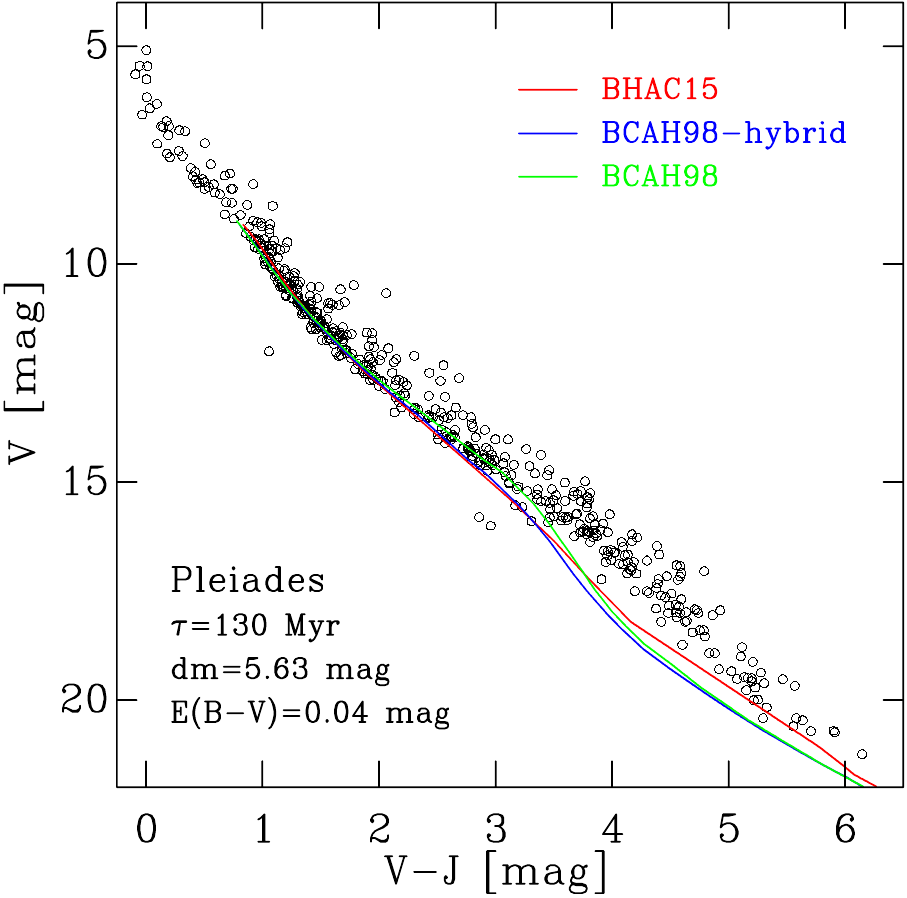}
\caption[]{$V, V-J$ CMD of the Pleiades. Overlaid are different
  `generations' of the Baraffe et al. model isochrones with ages of
  130\,Myr, each of which has been reddened by the equivalent of
  $E(B-V)=0.04\,\rm{mag}$ and shifted vertically by a distance modulus
  $dm=5.63\,\rm{mag}$. The original \cite{Baraffe98} models (with a
  solar-calibrated mixing-length parameter $\alpha=1.9$) coupled with
  the NextGen atmospheric models are shown as the green line. The same
  interior models, but coupled with the BT-Settl atmospheric models
  (computed using the \citealp{Asplund09} abundances) are represented
  by the blue line. The latest \cite{Baraffe15} interior models
  coupled with the BT-Settl models (both computed using the
  \citealp{Caffau11} abundances) are denoted by the red line. It is
  clear that whilst the discrepancy between the models and the data
  has decreased with time, there are still unresolved problems in the
  optical and/or near-IR bandpasses at low $T_{\rm{eff}}$.}
\label{fig:pleiades_cmd}
\end{figure}

In \cite{Baraffe15} the authors suggest that the new BHAC15 models now
match the shape of the observed locus, in the sense that the optical
colours for M dwarfs are no longer too blue. In
Fig.~\ref{fig:pleiades_cmd} we show the $V, V-J$ CMD of the Pleiades
and overlay 130\,Myr model isochrones from different `generations' of
the Baraffe et al. models (assuming the same distance modulus and
reddening as noted in the previous paragraph). From
Fig.~\ref{fig:pleiades_cmd} it is clear that whilst progress has been
made since the original \cite{Baraffe98} models were first published,
there are still unresolved problems with the latest BHAC15 in the
optical and/or near-IR bandpasses at low $T_{\rm{eff}}$. Repeating the
binary analysis of \cite{Bell12} in which we compared the theoretical
system magnitudes predicted by the theoretical models with the
measured system magnitudes in the $VI_{\rm{c}}JHK_{\rm{s}}$
bandpasses, we find that, for the BHAC15 models, the observed spread
in the optical bandpasses is smaller than for the other models,
however it is still appreciably larger than in the near-IR
bandpasses. Thus despite the fact that the discrepancy between
observed loci and the models is smaller when using the BHAC15 models
(compared with previous `generations'), it is still necessary to
calculate additional empirical corrections to the theoretical
BC-$T_{\rm{eff}}$ relations before using these to derive absolute ages
for young stellar populations using CMDs.

\section{Isochronal ages for young, nearby groups}
\label{isochronal_ages_for_young_clusters_associations}

We use the maximum-likelihood $\tau^{2}$ fitting statistic of
\cite{Naylor06} and \cite{Naylor09} to derive ages from the $M_{V},
V-J$ CMDs of our sample of young moving groups. The $\tau^{2}$ fitting
statistic can be viewed as a generalisation of the $\chi^{2}$
statistic to two dimensions in which both the model isochrone and the
uncertainties for each datapoint are two-dimensional
distributions. Not only does the use of the $\tau^{2}$ fitting
statistic remove the issue of objectivity introduced through `by-eye'
fitting of model isochrones, but it also allows us to include the
effects of binarity in our model distribution, provides reliable
uncertainties on the derived parameters and allows us to test whether
the model provides a good fit to the data. Similarly to $\chi^{2}$,
the best-fit model is found by minimising $\tau^{2}$.

\subsection{Creating the model distributions}
\label{creating_the_model_distribution}

The two-dimensional model distributions are generated using a Monte
Carlo method to simulate $10^{6}$ stars over a given mass range. To
populate our models we adopt the canonical broken power law mass
function of \cite*{Dabringhausen08}. An important feature of our
fitting method is that we include the effects of binary stars, thereby
modifying the model isochrone at a given age from a curve in CMD space
to a two-dimensional probability distribution. To include binary stars
in our fits we assume a binary fraction of 50 per cent and a uniform
secondary distribution ranging from zero to the mass of the primary.

Our sample of member stars ranges from early B-type stars to late
M-type dwarfs, and given that across this range there is an obvious
trend for the observed binary fraction to decrease as a function of
decreasing spectral type (or mass; see e.g. \citealp{Duchene13}), our
choice of a uniform binary fraction of 50 per cent for all spectral
types is somewhat idealised. The effect of varying the adopted binary
fraction in model distributions has previously been investigated by
\cite{Naylor06} who demonstrated that even at unrealistic values
(e.g. 80 per cent across all spectral types) the best-fit age is
affected at the $< 10$ per cent level. The model grids we calculate
are dense and cover a large age range [$\mathrm{log(age)}=6.0-10.0$ in
  steps of $\Delta \mathrm{log(age)}=0.01\,\rm{dex}$], and so given
this insensitivity to the adopted binary fraction, we adopt a uniform
fraction of 50 per cent for the ease of computing them.

We note that due to the processes involved in creating the
semi-empirical BC-$T_{\rm{eff}}$ relations (see
Section~\ref{the_models}), the lower $T_{\rm{eff}}$ limit of the BCs
may be reached before the lower mass limit of the interior
models. Effectively this means that the semi-empirical pre-MS model
isochrones only extend down to masses of $\simeq
0.1\,\rm{M_{\odot}}$. As a result, simulated secondary stars with
masses corresponding to $T_{\rm{eff}}$ values below this
$T_{\rm{eff}}$ threshold are assumed to make a negligible contribution
to the system flux i.e. this is equivalent to placing the binary on
the single-star sequence. Essentially, this means that we begin to
lose the binary population before the single-star population in our
two-dimensional model distributions and results in a `binary wedge' of
zero probability between the binary and single-star sequences at low
masses (see e.g. Fig.~\ref{fig:ab_dor_cmd}).

\subsection{Fitting the CMDs}
\label{fitting_the_cmds}

Whilst we have made every effort to include only high-probability
($\geq 90$ per cent) candidate members in our sample of young moving
groups, it is likely that a fraction of these stars are in fact
non-members. We have further improved the $\tau^{2}$ fitting statistic
since the modifications of \cite{Naylor09} and the reader is referred
to Appendix~\ref{tau2_non_members} in which we introduce an updated
method (cf. \citealp{Bell13}) to deal with non-member contamination by
assuming a uniform non-member distribution. We choose such a
conceptual framework, in part because it is better suited to the
problem in hand, but also because it allows us to calculate a
goodness-of-fit parameter, a step which was missing from our earlier
soft-clipping technique used in \cite{Bell13}.

\begin{figure*}
\centering
\includegraphics[height=10.5cm]{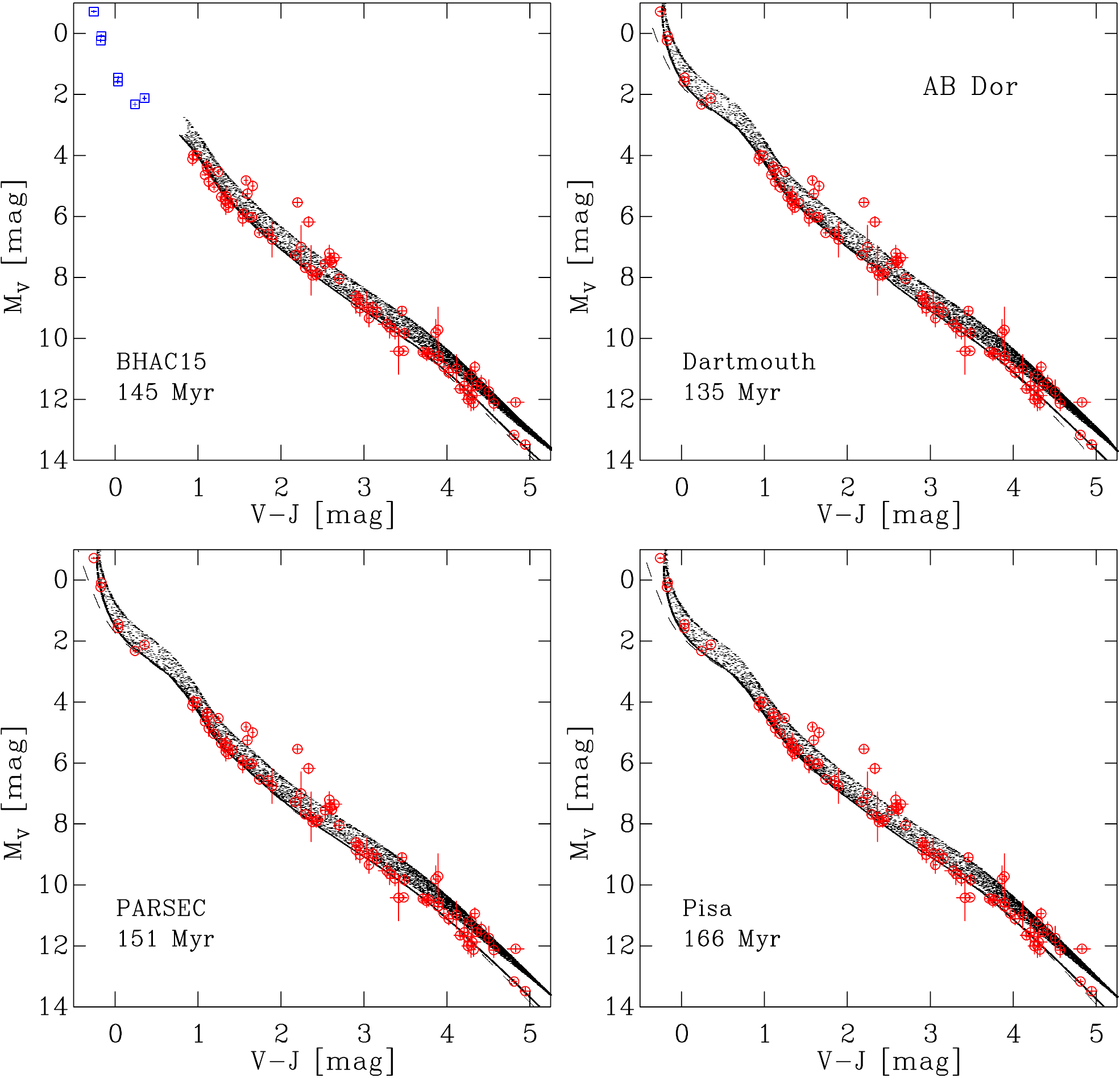}
\caption[]{Best-fitting $M_{V}, V-J$ CMDs of the AB Dor moving
  group. The red circles represent fitted data, whereas the blue
  squares denote objects which are removed prior to fitting (see
  Section~\ref{fitting_the_cmds} for details). \textbf{Top left:}
  BHAC15. \textbf{Top right:} Dartmouth. \textbf{Bottom left:}
  PARSEC. \textbf{Bottom right:} Pisa. The dashed line in each panel
  represents the ZAMS relation for that specific set of model
  isochrones.}
\label{fig:ab_dor_cmd}
\end{figure*}

Prior to deriving ages from the $M_{V}, V-J$ CMDs for our sample of
young groups, we must first address whether we need to account for
local levels of interstellar extinction and reddening in the solar
neighbourhood. All of our member stars lie within the so-called Local
Bubble ($\lesssim 100\,\rm{pc}$). A recent study by \cite{Reis11}
performed an analysis of the interstellar reddening in the Local
Bubble using Str{\"o}mgren photometry and provided estimates of how
$E(b-y)$ varies as a function of distance. To calculate how this local
reddening could affect the colours and magnitudes of our member stars,
we assume that $A_{V}=4.3 \times E(b-y)$ and $A_{J}/A_{V}=0.282$ (see
\citealp{Glaspey71} and \citealp{Rieke85} respectively) to infer that
the typical levels of extinction and reddening are $A_{V} \simeq
0.03\,\rm{mag}$ and $E(V-J) \simeq 0.02\,\rm{mag}$
respectively. Compared to the combined photometric and distance
uncertainties, such levels of extinction and reddening are
insignificant and therefore we can leave the semi-empirical models in
the absolute magnitude-intrinsic colour plane to derive ages from the
$M_{V}, V-J$ CMDs.

To illustrate our fitting procedure we will use the AB Dor moving
group as an example, for which the best-fitting $M_{V}, V-J$ CMDs are
shown in Fig.~\ref{fig:ab_dor_cmd}. Prior membership probabilities for
the individual stars are taken from the analyses of
\cite{Malo13,Malo14a,Malo14b}, and for this dataset the mean prior
probability of a given star being a member of the cluster $P(M_{\rm
  c})$ is greater than 0.99 (see
Appendix~\ref{formalism_and_implementation} for an explanation of how
the $\tau^{2}$ fitting statistic can be used to include prior
membership probabilities for individual sources). Therefore, with 89
stars in our catalogue, this would then imply that, at most, one
object does not originate from the model sequence. Note that for
member stars in our sample of young moving groups which are not
included in the analyses of \citeauthor{Malo13} we adopt a uniform
prior probability of 0.9.

\begin{figure}
\centering
\includegraphics[width=\columnwidth]{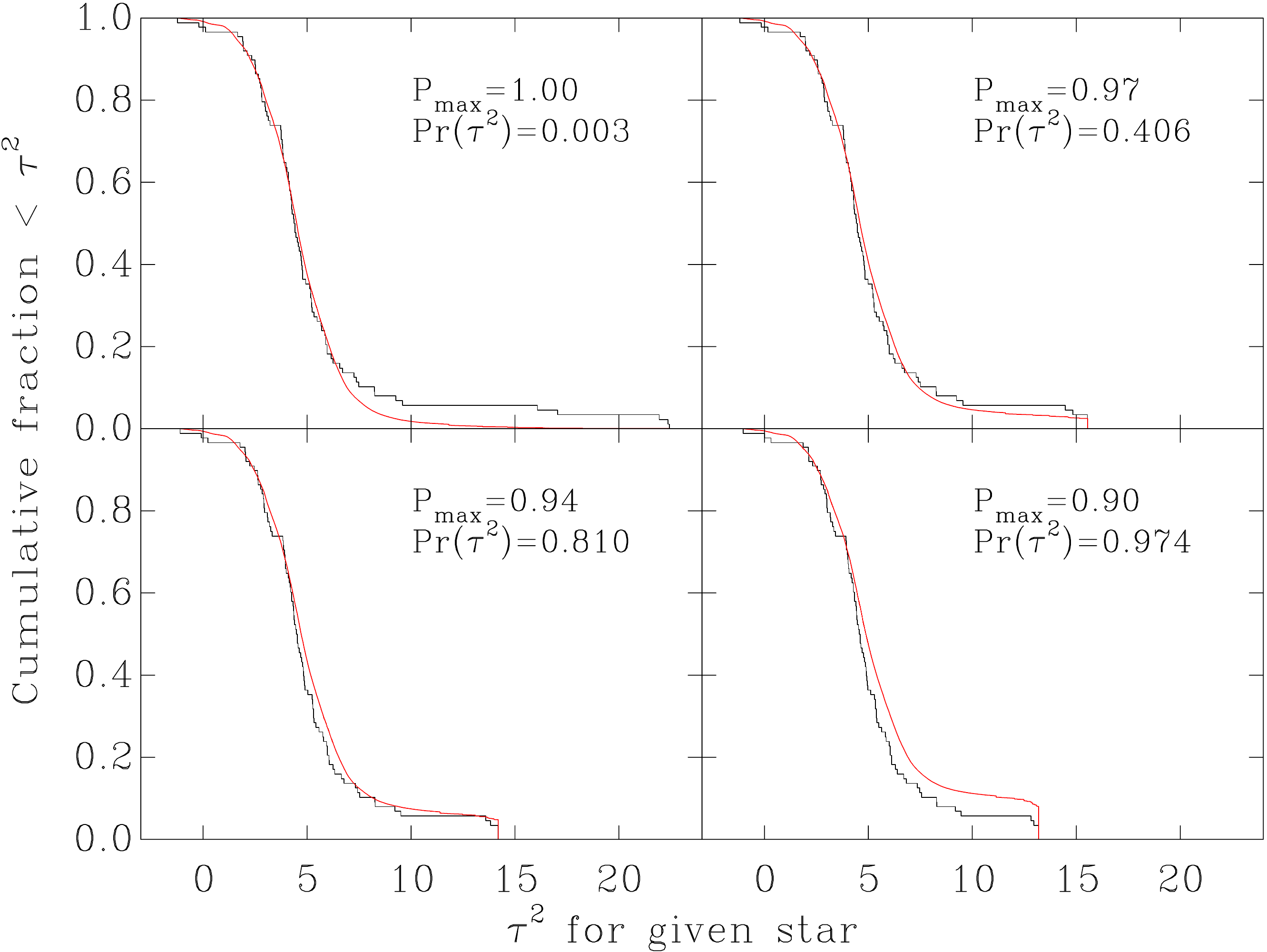}
\caption[]{Comparing the real and expected $\tau^{2}$ distributions
  (using the Dartmouth models) for the AB Dor moving group. The black
  histograms show the fraction of stars with individual $\tau^2$
  values less than a given value in the fit to the data. The red
  curves represent the expected distributions of the probability of
  obtaining a given $\tau^{2}$ calculated for the simulated
  observation (using the best-fit model). Each panel is for a
  different value of the maximum allowed prior membership probability
  $P_{\rm max}$ (see text).}
\label{fig:distrib}
\end{figure}

An examination of Fig.~\ref{fig:ab_dor_cmd}, however, suggests that
such a low non-member fraction is unlikely. There is a group of four
objects which lie significantly above the equal-mass binary envelope
of the model distribution between $1.5 \lesssim V-J \lesssim
2.5\,\rm{mag}$, the brightest of which is brighter than an equal-mass
quintuple. Furthermore, there are several additional objects which,
even allowing for the uncertainties in $M_{V}$, also lie above the
upper envelope of the best-fit model. It is worth noting that none of
these objects are single stars, but are either resolved components of
binary systems or unresolved binaries, all of which have separations
of $< 4\,\rm{arcsec}$, and therefore the effects of orbital motion
could conceivably affect the parallax solutions of
\cite{vanLeeuwen07}. Note that there is also a faint object (GJ~393)
at $M_V \simeq 10.4\,\rm{mag}$, $V-J \simeq 3.5\,\rm{mag}$, which is a
single star with a well-constrained distance, that lies $\simeq
0.5\,\rm{mag}$ below the single-star sequence. Such objects therefore
imply that the probability a given star is not modelled by our
sequence is greater than the 0.01 probability suggested by the raw
values of $P(M_{\rm c})$. This conclusion is justified by examining
the distribution of the $\tau^{2}$ values for each individual
datapoint. The top left panel of Fig.~\ref{fig:distrib} shows the
cumulative distribution for these values, along with the predicted
distribution calculated from the same 1000 simulated clusters (in this
case based on the Dartmouth model distributions) used to calculate the
goodness-of-fit parameter [$\Pr(\tau^{2})$; see
  Appendix~\ref{new_tau2_method}]. There is an obvious tail of points
with values much higher than the prediction from the simulations.

To overcome this problem we therefore set a maximum prior membership
probability ($P_{\rm max}$) and multiply all the prior membership
probabilities by this factor. We assessed the correct value for
$P_{\rm max}$ by adjusting it until the total of the prior and
posterior memberships roughly matched. This process is illustrated in
Fig.~\ref{fig:distrib}. The drop at the end of the distribution is the
stars which are apparent non-members, and this is fairly closely
matched at $P_{\rm max}=0.94$ which is the value where the total prior
and posterior memberships are best matched. For the AB Dor moving
group we find that, regardless of which set of semi-empirical pre-MS
models is adopted, the maximum prior membership probability is
$P_{\rm{max}} \geq 0.9$, and that this results in goodness-of-fit
values of $\Pr(\tau^{2}) \geq 0.5$ in all cases.

\begin{figure*}
\centering
\includegraphics[height=10.5cm]{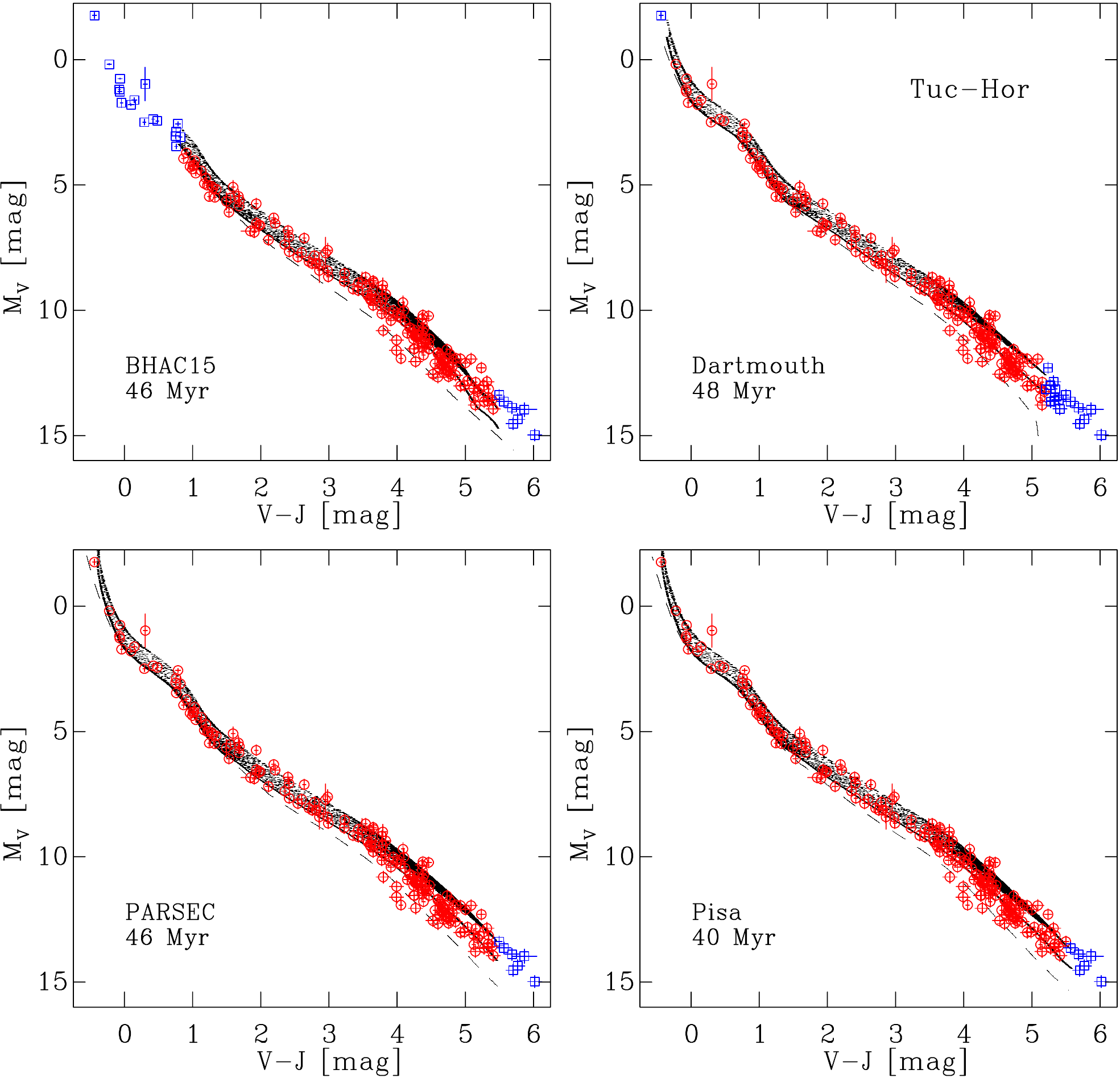}
\caption[]{Best-fitting $M_{V}, V-J$ CMDs of Tuc-Hor. The coloured
  symbols and dashed lines are the same as those in
  Fig.~\ref{fig:ab_dor_cmd}. \textbf{Top left:} BHAC15. \textbf{Top
    right:} Dartmouth. \textbf{Bottom left:} PARSEC. \textbf{Bottom
    right:} Pisa.}
\label{fig:tuc_hor_cmd}
\end{figure*}

An important feature to bear in mind when calculating reliable
best-fit ages and goodness-of-fit parameters is that there are certain
regions in CMD space which our grid of models [covering the age range
  $\mathrm{log(age)}=6.0-10.0$] simply do not occupy. Note that this
is compounded for the BHAC15 models which have an upper mass cut-off
of $1.4\,\rm{M_{\odot}}$, which means that any `handle' on the age
provided by higher mass stars (specifically those evolving between the
zero-age main-sequence [ZAMS] and the terminal-age main-sequence
[TAMS]) is effectively ignored by these models, which is not the case
for the other models which have upper mass cut-offs of $\geq
5\,\rm{M_{\odot}}$. Although we have tried to exclude known brown
dwarfs from our membership lists, there are still some stars which
occupy regions of the CMD not covered by the models (at both high- and
low-masses; see e.g. the blue squares in the upper right panel of
Fig.~\ref{fig:tuc_hor_cmd}), which must be removed from the fit before
calculating either of the aforementioned parameters.

\begin{figure*}
\centering
\includegraphics[height=10.5cm]{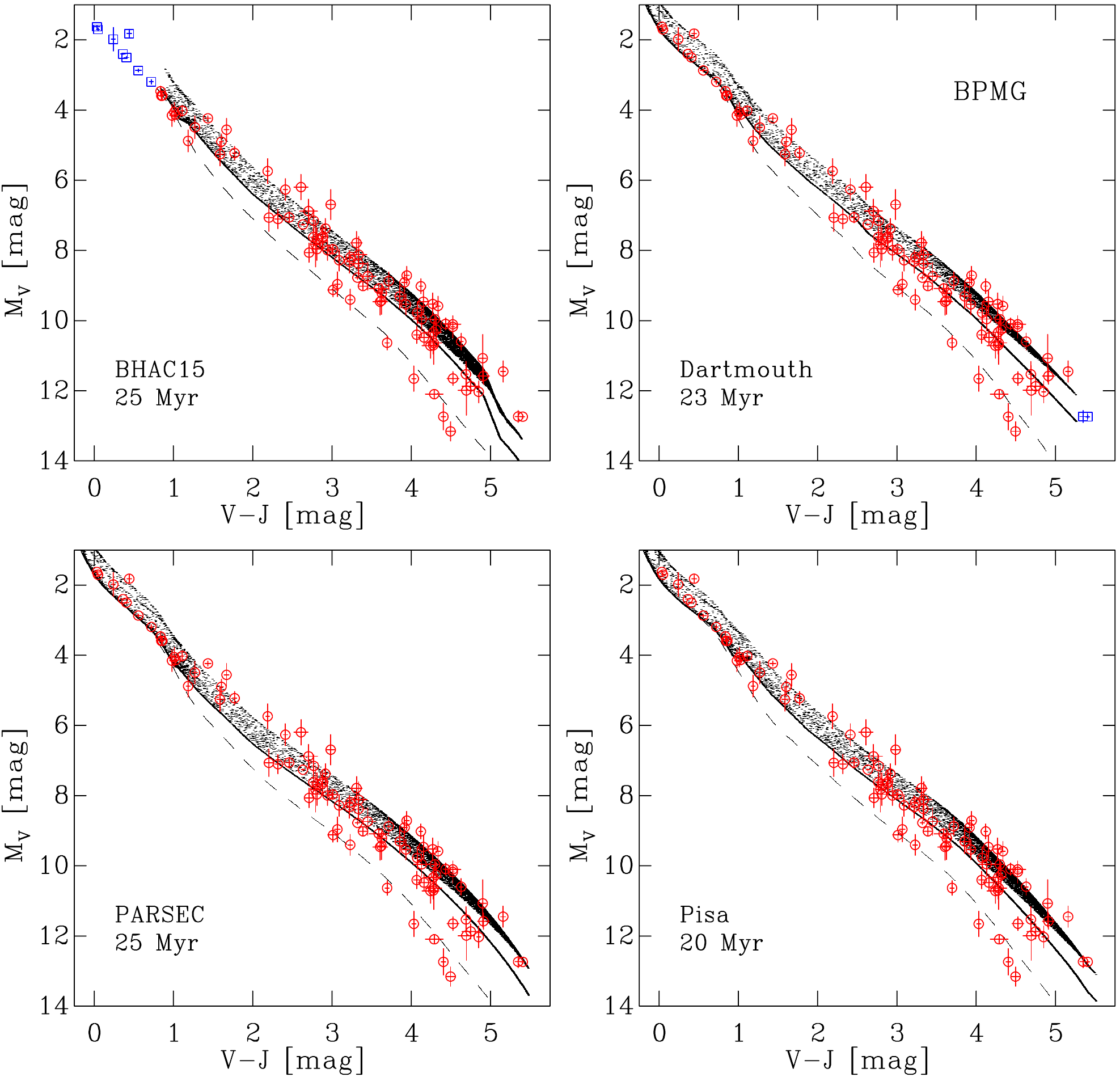}
\caption[]{Best-fitting $M_{V}, V-J$ CMDs of the BPMG. The coloured
  symbols and dashed lines are the same as those in
  Fig.~\ref{fig:ab_dor_cmd}. \textbf{Top left:} BHAC15. \textbf{Top
    right:} Dartmouth. \textbf{Bottom left:} PARSEC. \textbf{Bottom
    right:} Pisa.}
\label{fig:beta_pic_cmd}
\end{figure*}

A further consequence of our fitting technique is that our best-fit
model returns posterior membership probabilities. For a given group in
our sample we expect a negligible age spread (or equivalently
luminosity spread), and therefore we can use the posteriors to
identify stars which appear to be non-members (based solely on CMD
position) in an effort to further refine the membership lists of these
young moving groups. Such a technique is far from unassailable. For
example, we need only think of the obscuring effects of edge-on discs
or uncertainties in the distances to objects (see e.g. TWA~9 in
Section~\ref{data:distances}) which will act to shift stars away from
the observed locus of the group. Identification of such stars,
however, represents a sensible `first-order sanity check' which should
be used in combination with other metrics (e.g. Li equivalent widths,
radial velocities, etc.) to establish whether a given star is a
high-probability candidate member of a given moving group. For
example, the analyses of \citeauthor{Malo13} have identified several
high-probability candidate members in the BPMG, however an examination
of Fig.~\ref{fig:beta_pic_cmd} clearly shows that the inclusion of
these stars results in a luminosity spread of $\simeq 3\,\rm{mag}$ in
$M_{V}$ at a colour of $V-J \simeq 4.5\,\rm{mag}$. Such a spread,
given its age of $\gtrsim 20\,\rm{Myr}$, is incomprehensibly vast and
the fact that the faintest stars in this sample lie below the computed
ZAMS lends confidence to our assertion that our membership list still
contains likely non-members. This is also borne out by the calculated
posterior probabilities for these stars which are all less than 0.1
i.e. indicating that these are likely non-members.

\begin{figure*}
\centering
\includegraphics[height=10.5cm]{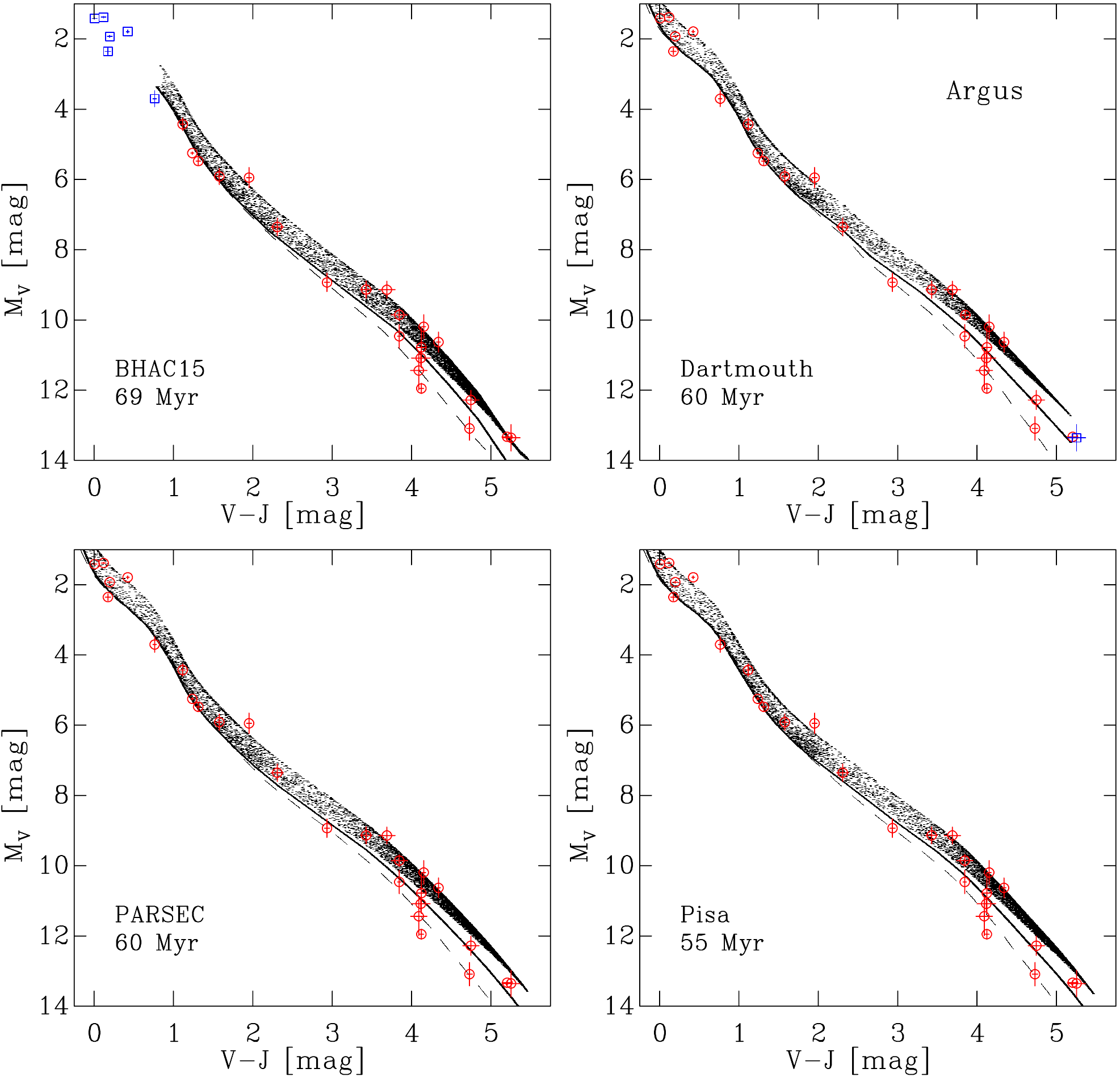}
\caption[]{Best-fitting $M_{V}, V-J$ CMDs of Argus. The coloured
  symbols and dashed lines are the same as those in
  Fig.~\ref{fig:ab_dor_cmd}. \textbf{Top left:} BHAC15. \textbf{Top
    right:} Dartmouth. \textbf{Bottom left:} PARSEC. \textbf{Bottom
    right:} Pisa.}
\label{fig:argus_cmd}
\end{figure*}

\begin{figure*}
\centering
\includegraphics[height=10.5cm]{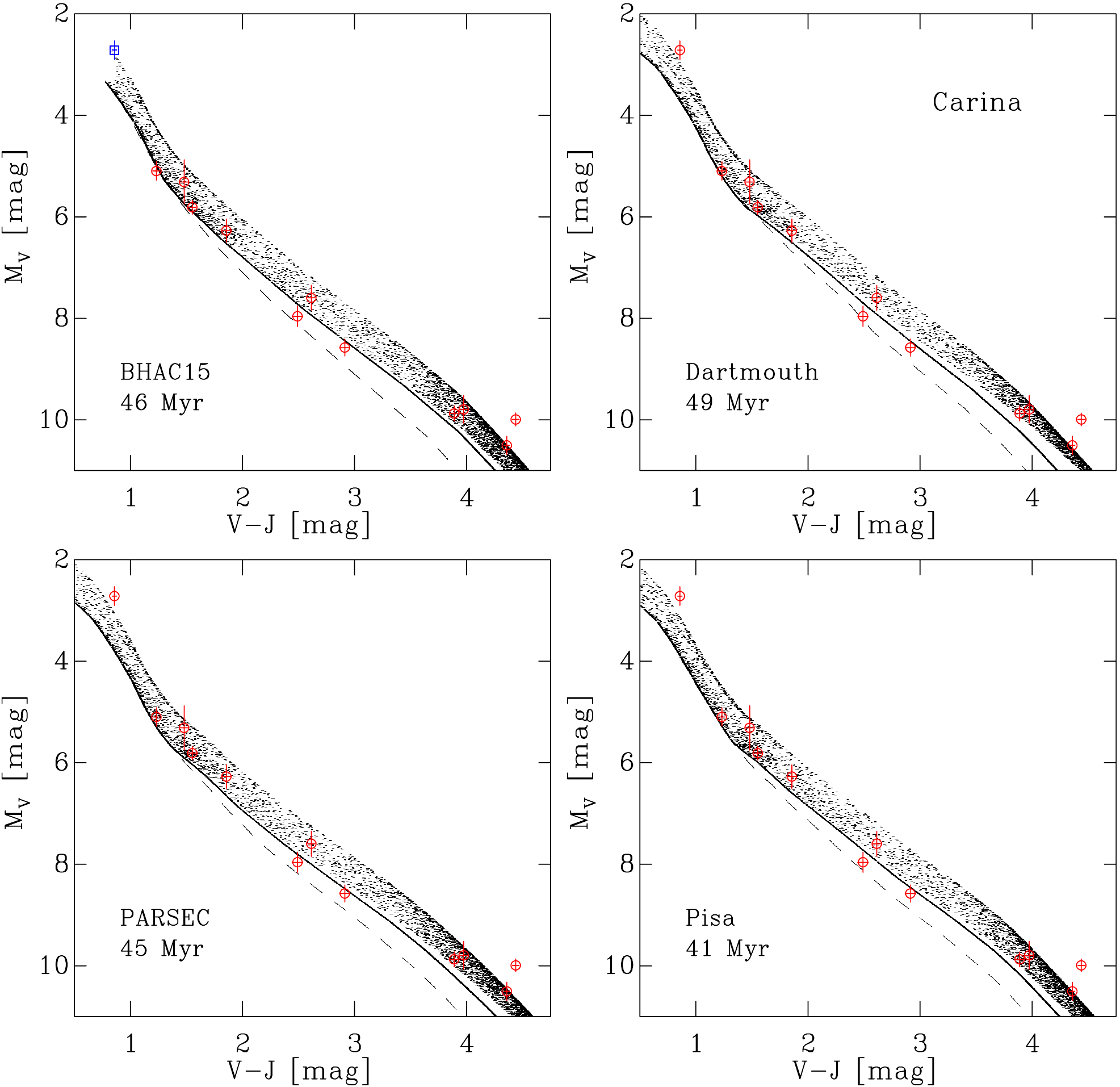}
\caption[]{Best-fitting $M_{V}, V-J$ CMDs of Carina. The coloured
  symbols and dashed lines are the same as those in
  Fig.~\ref{fig:ab_dor_cmd}. \textbf{Top left:} BHAC15. \textbf{Top
    right:} Dartmouth. \textbf{Bottom left:} PARSEC. \textbf{Bottom
    right:} Pisa.}
\label{fig:carina_cmd}
\end{figure*}

\begin{figure*}
\centering
\includegraphics[height=10.5cm]{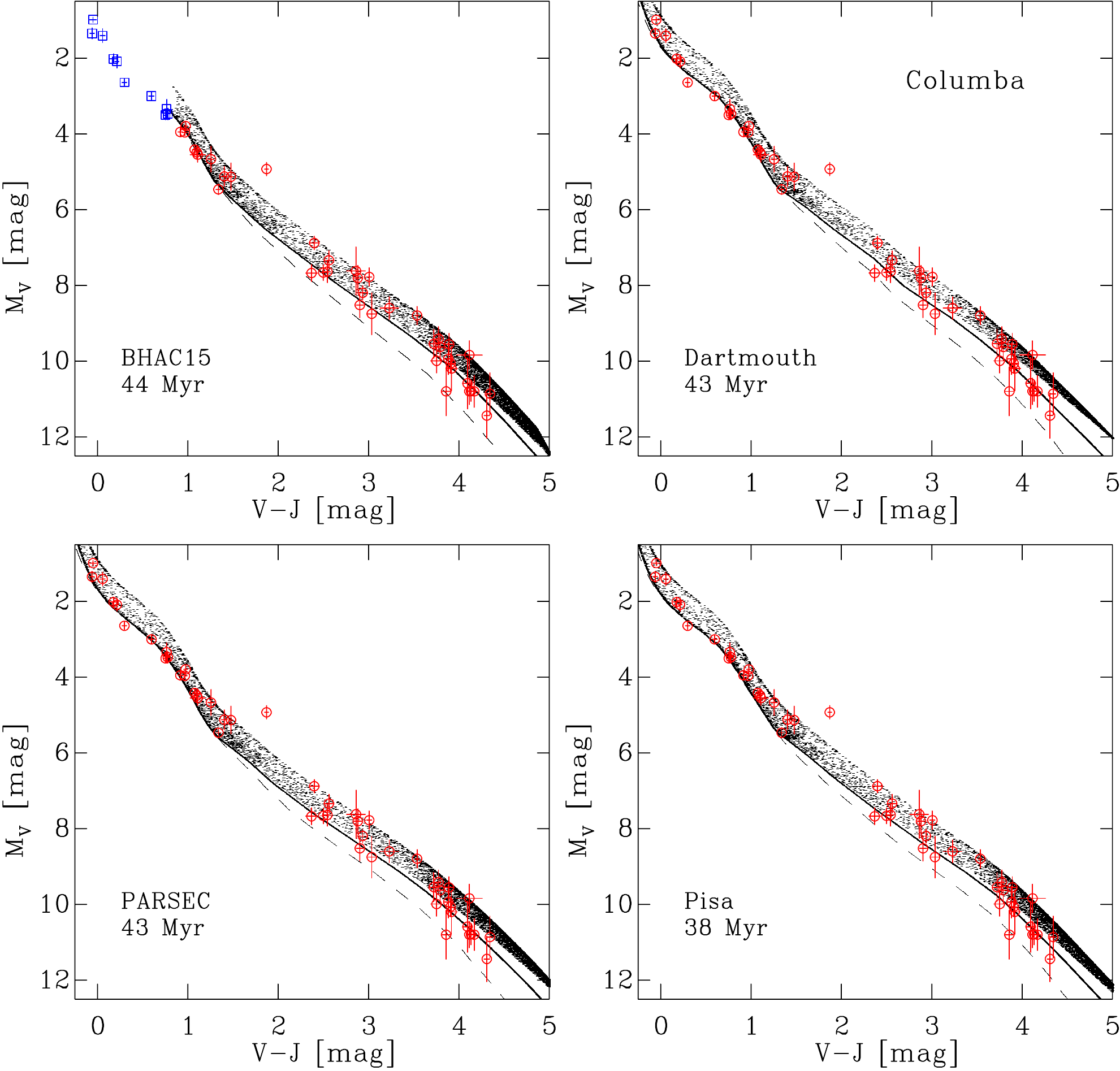}
\caption[]{Best-fitting $M_{V}, V-J$ CMDs of Columba. The coloured
  symbols and dashed lines are the same as those in
  Fig.~\ref{fig:ab_dor_cmd}. \textbf{Top left:} BHAC15. \textbf{Top
    right:} Dartmouth. \textbf{Bottom left:} PARSEC. \textbf{Bottom
    right:} Pisa.}
\label{fig:columba_cmd}
\end{figure*}

\begin{figure*}
\centering
\includegraphics[height=10.5cm]{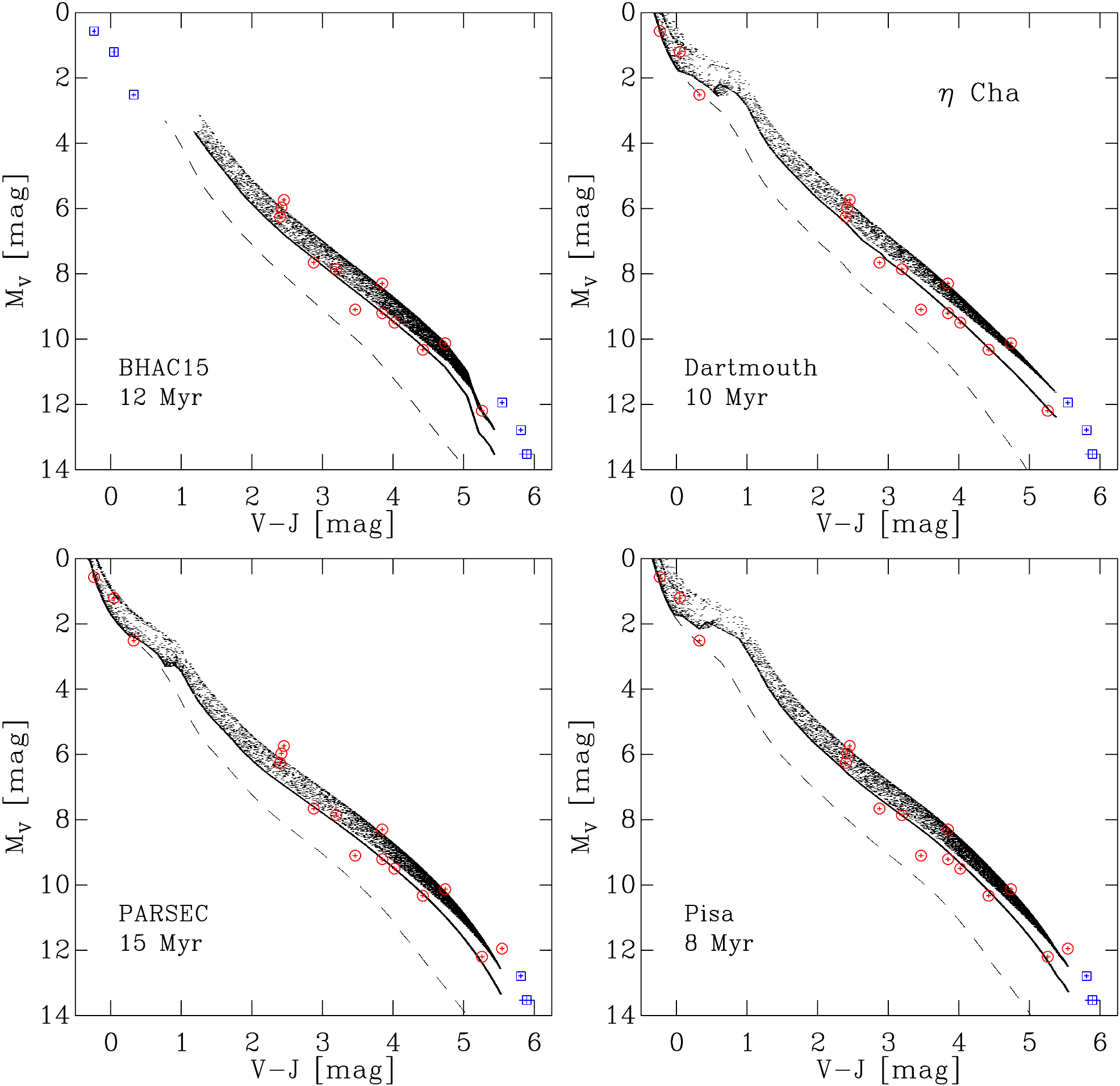}
\caption[]{Best-fitting $M_{V}, V-J$ CMDs of $\eta$ Cha. The coloured
  symbols and dashed lines are the same as those in
  Fig.~\ref{fig:ab_dor_cmd}. \textbf{Top left:} BHAC15. \textbf{Top
    right:} Dartmouth. \textbf{Bottom left:} PARSEC. \textbf{Bottom
    right:} Pisa.}
\label{fig:eta_cha_cmd}
\end{figure*}

\begin{figure*}
\centering
\includegraphics[height=10.5cm]{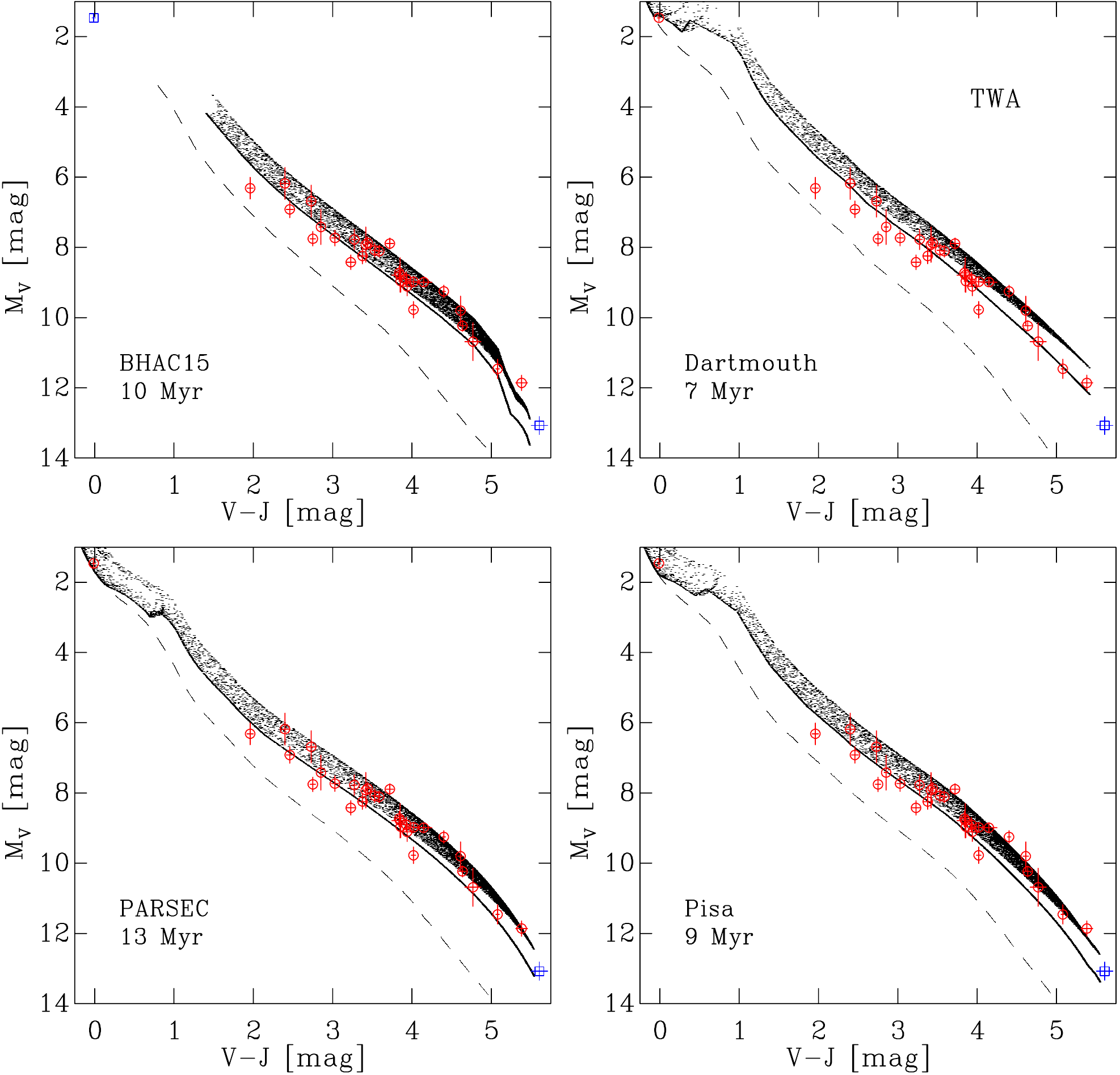}
\caption[]{Best-fitting $M_{V}, V-J$ CMDs of TWA. The coloured symbols
  and dashed lines are the same as those in
  Fig.~\ref{fig:ab_dor_cmd}. \textbf{Top left:} BHAC15. \textbf{Top
    right:} Dartmouth. \textbf{Bottom left:} PARSEC. \textbf{Bottom
    right:} Pisa.}
\label{fig:tw_hydrae_cmd}
\end{figure*}

\begin{figure*}
\centering
\includegraphics[height=10.5cm]{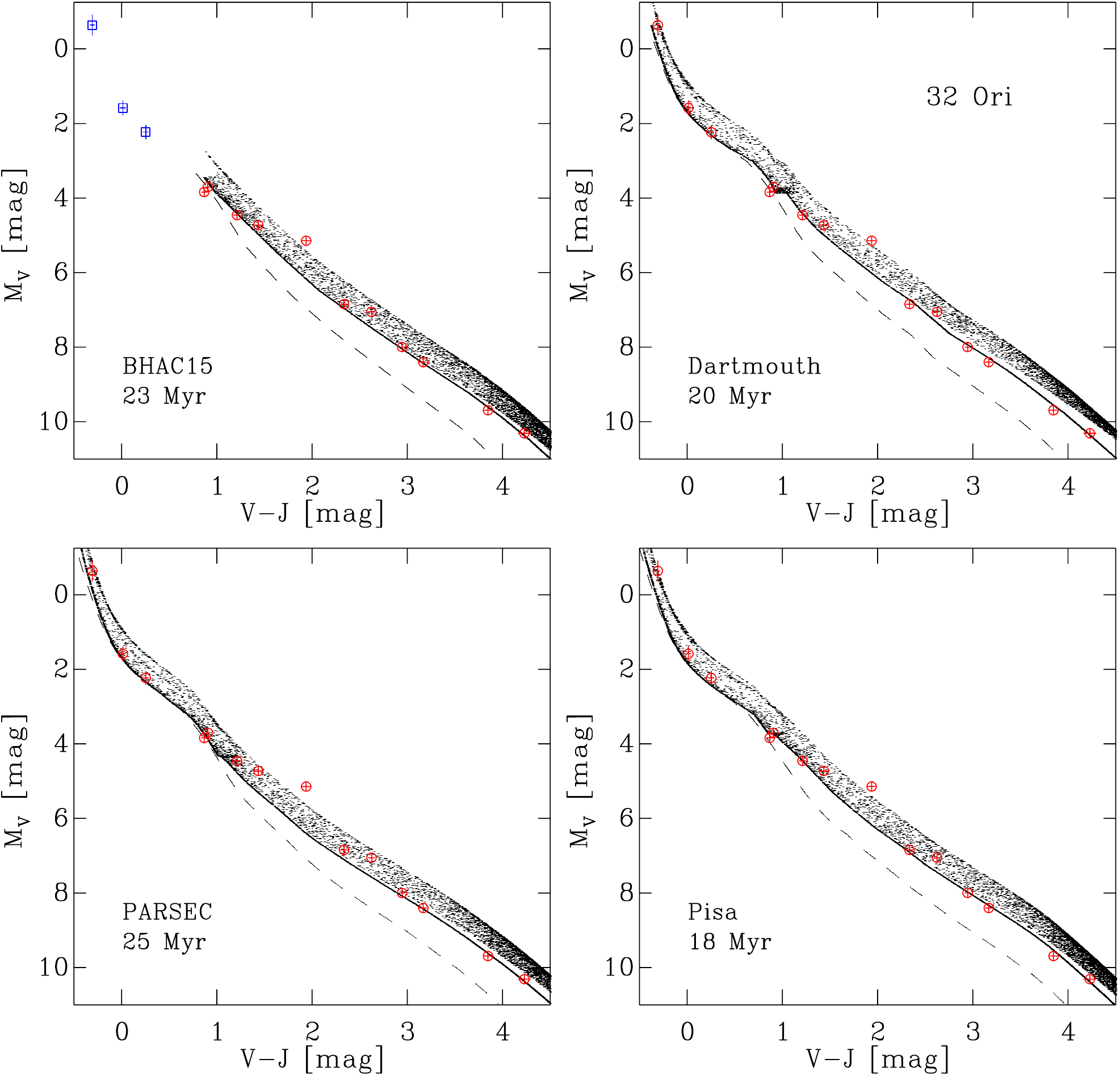}
\caption[]{Best-fitting $M_{V}, V-J$ CMDs of 32 Ori. The coloured
  symbols and dashed lines are the same as those in
  Fig.~\ref{fig:ab_dor_cmd}. \textbf{Top left:} BHAC15. \textbf{Top
    right:} Dartmouth. \textbf{Bottom left:} PARSEC. \textbf{Bottom
    right:} Pisa.}
\label{fig:32_ori_cmd}
\end{figure*}

Figs.~\ref{fig:argus_cmd}--\ref{fig:32_ori_cmd} show the best-fitting
$M_{V}, V-J$ CMDs for the remainder of our sample of young groups. Of
the remaining groups, we find that, except for Argus, the BPMG and
$\eta$ Cha, the maximum prior membership probability is $P_{\rm{max}}
\gtrsim 0.9$ and that the resultant goodness-of-fit values
$\Pr(\tau^{2})$ are $\gtrsim 0.5$ for all four sets of semi-empirical
pre-MS models. It is likely that the lower $P_{\rm{max}}$ values
required for the BPMG and $\eta$ Cha (of between 0.7 and 0.8) are a
result of contamination by non-members in the former (see above) and
astrophysical phenomena affecting the CMD positions of $\simeq 10$ per
cent of the members in the latter (see
Section~\ref{discussion_eta_cha}) in conjunction with our adoption of
a well-constrained distance ($< 2$ per cent uncertainty) for all
member stars. Whilst the derived $P_{\rm{max}}$ values for Argus are
all $\gtrsim 0.85$, the corresponding goodness-of-fit values
$\Pr(\tau^{2})$ are $\lesssim 0.2$ for all of the adopted
models. Again, it is likely that this is suggestive of a $\simeq
10-15$ per cent level contamination in our membership list for this
association (see Section~\ref{discussion_arga}).

Table~\ref{tau2_results} shows the best-fit age for each group
according to the four sets of semi-empirical models, in addition to
our final adopted age. For the final age we adopt the average of i)
the median, ii) the Chauvenet clipped mean \citep{Bevington92}, and
iii) the probit mean \citep{Lutz80} of the individual best-fit
ages. The quoted uncertainties on the final ages represent the
statistical and systematic uncertainties added in quadrature. Given
the asymmetric statistical uncertainties on several of the individual
ages shown in Table~\ref{tau2_results}, we calculate both an upper and
lower statistical uncertainty by taking the median of the four
individual upper and lower statistical uncertainties. Our estimate of
the systematic uncertainty arising from the use of different model
isochrones is based on the average of i) the 68 per cent confidence
levels and ii) the probit standard deviation \citep{Lutz80} of the
individual best-fit ages.

\begin{table*}
\caption[]{Ages for the young groups studied in this paper. Individual
  ages are shown for each set of semi-empirical pre-MS model
  isochroness and have been derived using the $\tau^{2}$ fitting
  statistic for which we have set $P_{\rm{max}}$ so that the total
  prior and posterior memberships are roughly equal (see
  Section~\ref{fitting_the_cmds} for details). The penultimate row
  lists our final adopted age for each group on which the quoted
  uncertainties represent the statistical and systematic uncertainties
  added in quadrature. The final row shows literature LDB ages for the
  BPMG and Tuc-Hor (see Section~\ref{comparison_to_ldb_ages} for
  references) which highlights the consistency between the two age
  diagnostics.}
\begin{tabular}{l c c c c c c c c c}
  \hline
Model   &   \multicolumn{9}{c}{Group age (Myr)}\\
              &   AB Dor   &   Argus$^{a}$   &   BPMG   &    Carina   &   Columba   &   $\eta$ Cha   &   Tuc-Hor   &   TWA   &   32 Ori\\
\hline
BHAC15   &   $145^{+889}_{-5}$   &   $69^{+19}_{-8}$   &   $25\pm1$   &   $46^{+16}_{-3}$   &   $44^{+8}_{-3}$   &   $12\pm1$   &   $46^{+1}_{-2}$   &   $10\pm1$   &   $23^{+3}_{-2}$\\
Dartmouth   &   $135^{+15}_{-9}$   &   $60^{+64}_{-3}$   &   $23\pm1$   &   $49^{+13}_{-5}$   &   $43^{+8}_{-3}$   &   $10\pm1$   &   $48^{+1}_{-2}$   &   $7^{+2}_{-1}$   &   $20^{+5}_{-1}$\\
PARSEC   &   $151^{+24}_{-18}$   &   $60^{+268}$   &   $25^{+1}_{-2}$   &   $45^{+7}_{-12}$   &   $43^{+3}_{-4}$   &   $14\pm1$   &   $46\pm1$   &   $13\pm1$   &   $25^{+2}_{-5}$\\
Pisa       &   $166^{+74}_{-23}$   &   $55^{+235}_{-5}$   &   $20^{+1}_{-2}$   &   $41^{+8}_{-6}$   &   $38\pm3$   &   $8\pm1$   &   $40^{+1}_{-2}$   &   $9\pm1$   &   $18\pm1$\\
\hline
\textbf{Adopted}   &   $\mathbf{149^{+51}_{-19}}$   &   \textbf{--}   &   \textbf{24$\pm$3}   &   $\mathbf{45^{+11}_{-7}}$   &   $\mathbf{42^{+6}_{-4}}$   &   \textbf{11$\pm$3}   &   \textbf{45$\pm$4}   &      \textbf{10$\pm$3}   &   $\mathbf{22^{+4}_{-3}}$   \\
\hline
LDB age   &   --   &   --   &   $24\pm5$   &   --   &   --   &   --   &   $40\pm3$   &   --   &   --   \\
\hline
\end{tabular}

\vspace{1pt}
\begin{flushleft}
\textbf{Notes:}\\ $^{a}$We do not provide a final adopted age for
Argus as it remains unclear whether the stars in our list of members
are representative of a single population of coeval stars (see
Section~\ref{discussion_arga} for details). Note also that the
best-fit age for Argus using the PARSEC models only provides an upper
limit on the age of the association and hence we do not include a
lower age uncertainty in the table.
\end{flushleft}
\label{tau2_results}
\end{table*}

\section{Discussion}
\label{discussion}

In Section~\ref{fitting_the_cmds} we used semi-empirical models to
derive a self-consistent, absolute isochronal age scale for nine
young, nearby moving groups within 100\,pc of the Sun. To assess the
reliability of these ages, we must first compare these to what are
considered to be well-constrained ages for the same groups.

\subsection{Comparison to LDB ages}
\label{comparison_to_ldb_ages}

As discussed in the Introduction, LDB ages are arguably the most
reliable age diagnostic -- in terms of calculating absolute ages --
currently available for stellar populations with ages of between $\sim
20$ and 200\,Myr. Note that although LDB ages have been advocated as
our best hope of establishing a reliable age scale in this age range
(primarily due to the fact that LDB ages have been shown to be
relatively insensitive to variations in the physical inputs adopted in
the evolutionary models and the fact that between different models
there is excellent agreement; see e.g. \citealp{Soderblom14}), recent
studies have demonstrated that by not accounting for the effects of
starspots in the evolution of pre-MS stars, the LDB age scale may in
fact only be good to $\simeq 10-20$ per cent (see
e.g. \citealp{Jackson14b,Somers15}).

\cite{Binks14} and \cite{Malo14b} have both identified the LDB in the
BPMG, deriving ages of $21\pm4$ and $26\pm3\,\rm{Myr}$ respectively
(the difference in age stemming from the use of stellar evolutionary
models which include a magnetic field model in the latter
study). Despite this difference, the derived LDB ages are consistent
and for the purposes of comparing this to our isochronal age we adopt
a value of $24\pm5\,\rm{Myr}$. In addition to the available LDB age
for the BPMG, \cite{Kraus14} recently identified the LDB in Tuc-Hor,
for which we adopt an age of $40\pm3\,\rm{Myr}$. It is therefore
reassuring then to find that the isochronal ages we have calculated
for both the BPMG and Tuc-Hor are consistent with the derived LDB
ages. Note that our age for the BPMG also compares favourably to the
$22\pm3\,\rm{Myr}$ derived by \cite{Mamajek14} which was based on an
$M_{V}, B-V$ CMD analysis of the A-, F- and G-type members of the
group. Whilst such consistency lends confidence to our absolute age
scale for the young moving groups in the solar neighbourhood, there
are (currently) no other LDB ages for our sample with which to compare
the isochronal ages against. Hence, we must place the isochronal ages
for the seven remaining groups in context by comparing them to other
age-dating techniques adopted in the literature.

\subsection{AB Dor moving group}
\label{discussion_abdmg}

Literature ages for the AB Dor moving group vary from the relatively young ($\simeq 50-70\,\rm{Myr}$; see e.g. \citealp*{Zuckerman04b}; \citealp{Torres08,daSilva09}) to essentially coeval with the Pleiades ($\simeq 130\,\rm{Myr}$; see e.g. \citealp*{Luhman05c}). A more recent analysis by \cite{Barenfeld13} provided a strong constraint on the age of the group by identifying the main-sequence turn-on for AB Dor nucleus stars in the $M_{V}, V-K_{\rm{s}}$ CMD. They demonstrated that the late K-type stars have already settled onto the ZAMS, and used this to place a firm lower limit on the age of 110\,Myr. 

Interestingly, the recent discovery of a Li-rich M8 high-probability brown dwarf candidate member (2MASS J00192626+4614078; \citealp{Gagne14}) provides us with a strong upper limit on the age of the group of $196\,\rm{Myr}$ (Binks, private communication). Additional candidate members with spectral types in the range M4-M8 have been reported by \cite{Gagne15} and thus spectroscopic follow-up of these objects could place stronger constraints on the absolute age of the AB Dor moving group. Our isochronal age therefore lies almost exactly midway between the lower and upper age limits provided by the main-sequence turn-on and the lower edge of the tentative identification of the LDB.

From Table~\ref{tau2_results}, it is clear that the main `handle' on the age of the AB Dor moving group comes from the high-mass stars which are evolving between the ZAMS and TAMS. The BHAC15 models demonstrate that if these stars are omitted from the fit then, although the best-fit age is in general agreement with those from the other models, the associated upper uncertainty on the age is prohibitively large. The reason why the low-mass population of the group cannot provide a precise age is due simply to the fact that at ages of $\gtrsim 100\,\rm{Myr}$ the model isochrones occupy essentially the same position in CMD space (except for very-low-mass objects with $V-J \gtrsim 4.5\,\rm{mag}$ for which we have only a few stars in our AB Dor moving group membership list; see Fig.~\ref{fig:ab_dor_cmd}), resulting in a relatively flat $\tau^{2}$ space across a wide range of ages.

\subsection{Argus association}
\label{discussion_arga}

The kinematics of Argus are very similar to those of the nearby young cluster IC~2391 \citep{Torres03}, for which \cite*{Barrado04b} derived an LDB age of $50\pm5\,\rm{Myr}$. A similar age for Argus ($\sim40\,\rm{Myr}$) was proposed by \cite{Torres08} on the basis of Li equivalent widths and positions in the $V, V-I_{\rm{c}}$ CMD, and therefore our isochronal ages of $\sim 60\,\rm{Myr}$ disagree with the current literature age for this association. Furthermore, whilst the different sets of semi-empirical models tend to agree upon an age of $\sim 60\,\rm{Myr}$ for Argus, the uncertainties on these individual ages are extremely large and imply that based on our current membership list, it is not valid to assign a unique, unambiguous age to Argus (see Table~\ref{tau2_results}). We believe that the smaller uncertainties on the BHAC15 best-fit age stem from the upper-mass cut-off of these models which neglect any age information from the higher mass members. Excluding the BHAC15 models for the moment, the other ages in Table~\ref{tau2_results} clearly demonstrate that although there is a minimum $\tau^{2}$ value within the grid, the $\tau^{2}$ space is also relatively flat over a wide range of ages (see e.g. the PARSEC models which can only give an upper age limit of 328\,Myr).

Our membership list for Argus stems primarily from the Bayesian analyses of \cite{Malo13,Malo14a,Malo14b} and comparing it with the membership list of \citet[their Table~12]{Torres08} there are two notable differences i) only three stars in common appear in both lists (BW~Phe [HD~5578], HD~84075 and NY~Aps [HD~133813]) and ii) the median distance of members in our list is 32\,pc whereas for the \citeauthor{Torres08} list it is 96\,pc. Furthermore, looking also at the distribution of distances (from the nearest to the farthest) in these two membership lists, those stars in our list cover a range of $\simeq 65\,\rm{pc}$, whereas those in \cite{Torres08} cover a range of more than twice this at $\simeq 150\,\rm{pc}$ (cf. $\simeq 50\,\rm{pc}$ for the supposedly coeval Tuc-Hor). Given this spread in distances for the \citeauthor{Torres08} members, one can reasonably ask whether we expect these objects to have formed together and therefore be coeval in the first place.

Fig.~\ref{fig:argus_cmd} shows the best-fitting CMDs of Argus and there are two points worth mentioning. First, there is a group of 5 A-type stars (all of which were proposed as members by \citealp{Zuckerman11} and none of which are unresolved binaries [according to the Washington Visual Double Star Catalog; \citealp{Mason01}]), which we would expect to define the ZAMS of the association (as is the case with the other groups in our sample). None of these, however, actually lie on the ZAMS (4 are over-luminous and 1 is under-luminous with respect to the ZAMS), which suggests that these stars may not be coeval, but rather represent stars at different evolutionary stages. Second, of the K- and M-type stars in the association, a significant fraction of these appear to lie below the best-fitting model in all panels of Fig.~\ref{fig:argus_cmd}. Whilst these stars all appear to be young and active, as evidenced by a combination of high $R_{X} = L_{X}/L_{\rm{bol}}$ values and large H$\alpha$ equivalent widths (see e.g. \citealp*{Riaz06}), they all lack Li measurements which may help us discriminate between young active stars belonging to a given group and slightly older active ZAMS stars which do not (at least for the mid to late M-type objects). Given these ambiguities, we are therefore reluctant (at present) to assign a final, unambiguous age to Argus as it appears as though our membership list suffers from a high level of contamination, and hence it remains unclear whether this list represents a single, coeval population of stars, or even whether the association is in fact physical.

\subsection{Carina and Columba associations}
\label{discussion_cara_cola}

Carina is an extremely sparse association whose Galactic space motion is, to within the uncertainties, statistically indistinguishable from that of Columba (see e.g. \citealp{Malo14a}), although spatially the two associations are rather distinct. Both associations were identified by \cite{Torres08} who further demonstrated that they share a similar age of $\sim40\,\rm{Myr}$ through a combination of Li equivalent widths and positions in the $V, V-I_{\rm{c}}$ CMD. Our isochronal ages for both Carina and Columba are consistent with ages of $\sim 40\,\rm{Myr}$ and, to within the uncertainties, appear to be coeval (as well as share a common age with Tuc-Hor; see \citealp{Torres08}).

Fig.~\ref{fig:columba_cmd} demonstrates that our best-fit models appear to follow the observed locus of Columba (tracing both the upper and lower envelopes of the relatively well-populated association). For the much sparser Carina association, however, such agreement between the models and the data is not so obvious [see Fig.~\ref{fig:carina_cmd}; this is also reflected in the systematically lower $\Pr(\tau^{2})$ values for Carina compared to Columba]. Although we only have 12 stars in our membership list for Carina, there is a case to be made that there are two apparent outliers which appear to lie below the observed locus. These stars (namely 2MASS~J04082685-7844471 and 2MASS~J09032434-6348330) will act to `drag' the best-fit model to older ages, the result of which, is that 2MASS~J10140807-7636327 (the reddest star in our list) is assigned a posterior membership probability which indicates non-member status, despite occupying a position in CMD space which is commensurate with its binary status i.e. $\simeq 0.5\,\rm{mag}$ above the single-star sequence.

If we repeat our isochronal age analysis having removed 2MASS~J04082685-7844471 and 2MASS~J09032434-6348330, we find a slightly younger age for Carina of $\simeq 36\,\rm{Myr}$ (cf. 45\,Myr). It remains unclear whether the kinematic distance estimates for these two stars are erroneous or whether they are older than the other stars in our list. To better constrain the age of Carina, it is therefore clear that we have to determine whether these stars are genuine members of Carina or whether they are simply older active stars (akin to what we previously discussed in Argus; see Section~\ref{discussion_arga}). Given the spectral types of these two stars (early M-type) it is unclear whether further spectroscopic information (e.g. Li equivalent widths) would allow us to differentiate between these two options, and so this may be a case of having to wait until \emph{Gaia} provides us with the necessary kinematic information to unambiguously do so.

\subsection{$\eta$ Cha cluster}
\label{discussion_eta_cha}

The use of the LDB technique to derive ages is restricted to populations with ages of $\gtrsim 20\,\rm{Myr}$ and thus $\eta$ Cha is too young for the adoption of such a technique. There are, however, model-independent methods of deriving an age which can then be used as an independent diagnostic to compare against isochronal ages. One such method involves calculating the time of minimum separation between stellar groups in the past on the assumption that the groups share a common origin. As with other kinematic methods this approach has led to contradictory conclusions. For example, \cite*{Jilinski05} performed a kinematic traceback of the $\eta$ Cha and $\epsilon$ Cha clusters assuming a Galactic potential and found that the smallest separation of only a few pc was $\sim 7\,\rm{Myr}$ ago. This coevality of the two clusters was later refuted by \cite{Lawson09} who demonstrated that both clusters have significantly different ages based on a combination of CMD positions and surface gravity indicators. Furthermore, a more recent kinematic analysis by \cite{Murphy13} concluded that not only is $\eta$ Cha a few Myr older than $\epsilon$ Cha, but also that there is little evidence that the two clusters were appreciably closer than their current $\sim 30\,\rm{pc}$ in the past.

The literature isochronal ages for $\eta$ Cha imply an age of $5-8\,\rm{Myr}$ (e.g. \citealp{Luhman04}), however this becomes slightly older when the effects of binary stars are accounted for (cf. $7-9\,\rm{Myr}$; \citealp{Lyo04}). The fundamental parameters of the double-lined eclipsing binary RECX~8 (RS~Cha) are arguably some of the best constrained in our entire sample of member stars, with typical uncertainties of only a few per cent. Such precise measurements naturally provide stringent tests of stellar evolutionary models by forcing them to fit both components of the binary system at a given age (see e.g. \citealp*{Stassun14}). Simultaneous fitting of the masses, radii, $T_{\rm{eff}}$ and $L_{\rm{bol}}$ for both components by \cite{Alecian07} and \cite*{Gennaro12} imply an age of $\simeq 9\,\rm{Myr}$ for the RECX~8 system, which is consistent with our isochronal age of $\simeq 11\,\rm{Myr}$ for the cluster. Furthermore, Fig.~\ref{fig:eta_cha_cmd} shows that our semi-empirical pre-MS models provide a good fit to the data, tracing both the lower and upper envelope of the observed locus, except for the two outliers RECX~13 and RECX~15.

\cite{Luhman04} demonstrated that unlike the other members of $\eta$ Cha, which exhibit negligible levels of interstellar extinction, RECX~13 has a measured extinction of $A_{V}=0.4\,\rm{mag}$ [which corresponds to $E(V-J)=0.29\,\rm{mag}$ based on the $A_{J}/A_{V}=0.282$ relation of \citealp{Rieke85}; see Section~\ref{fitting_the_cmds}]. De-reddening RECX~13 using these values would place the star much closer to the single-star sequence of the model distribution. RECX~15, on the other hand, has a large IR excess at wavelengths of $\gtrsim 2\,\rm{\mu m}$ which is indicative of circumstellar material (see e.g. \citealp{Megeath05}). The inclination of this disc with respect to our line-of-sight remains ill-constrained, however high inclinations for discs around other $\eta$ Cha members have been reported ($\gtrsim 60^{\rm{o}}$; \citealp*{Lawson04}). If the disc around RECX~15 is so inclined, then we would expect significant dimming in the optical wavelengths as a result of observing the star through it's disc, and this could therefore explain why RECX~15 appears fainter than the other stars which represent the lower envelope of the cluster locus in CMD space.

\subsection{TWA}
\label{discussion_twa}

TWA was the first of the young moving groups in our sample to be identified in the literature (see e.g. \citealp{delaReza89,Kastner97}) and as such many age estimates are now available for this association. Several authors have attempted to use kinematic information to derive a model-independent age for the TWA, however, they are either contradictory or cover a prohibitively large range to provide a strong constraint on the age. For example, \cite{Mamajek05} noted that whilst the data are consistent with expansion, the corresponding expansion age of $20^{+25}_{-7}\,\rm{Myr}$ has such large associated uncertainties that it is of limited use.

More recent analyses by \cite*{Weinberger13} and \cite{Ducourant14}, which have measured a larger number of parallaxes and proper motions than in the \cite{Mamajek05} study, have come to completely opposite conclusions. \cite{Weinberger13} find that the space motions of TWA members are essentially parallel and do not indicate convergence at any time in the past 15\,Myr, whereas \cite{Ducourant14} find that a subset of 16 members occupied the smallest volume of space $\sim 7.5\,\rm{Myr}$ ago. Note that in the \cite{Ducourant14} study, of the 25 stars with reliable radial velocity and parallax information, the authors removed 9 stars from the sample (36 per cent) prior to deriving the traceback age as they demonstrated discrepant space motions. Omissions such as these again raise the question of subjectivity concerning the stars which one includes/excludes in a given sample.

Despite the lack of an unambiguous kinematic age for the TWA, several isochronal ages have been reported in the literature which suggest an age of $\sim 10\,\rm{Myr}$ (see e.g. \citealp{Webb99,Barrado06,Weinberger13}). Whilst our isochronal age is in excellent agreement with previous isochronal age estimates, Fig.~\ref{fig:tw_hydrae_cmd} shows that the Dartmouth models (in particular) do not provide a good fit to the data. Whilst it is apparent that these models have an enlarged `binary wedge' when compared to the other models (presumably stemming from the lower-mass cut-off in the interior models and the slightly different $T_{\rm{eff}}$ scale) which could play a role in this discrepancy, it is also notable that there is a mismatch between the observed slope of the locus and that of the best-fitting model. Furthermore, even the other models appear to lie slightly above the lower envelope of the TWA locus. Hence, it is possible that our isochronal age could be underestimated and that a more representative age would be closer to $\sim 15\,\rm{Myr}$.

Even when using models which include the effects of binary stars, it is clear that we are unable to fit both the lower and upper envelopes of the observed locus in the CMD of the TWA. One reason for this increased luminosity spread could be due to obscuration effects arising from discs around the stars which define the lower envelope (namely TWA~6, TWA~9B, TWA~21 and 2MASS J10252092-4241539) which, depending on their orientation, may act to make them appear dimmer in the CMD (see e.g. RECX~15 in Section~\ref{discussion_eta_cha}). A recent analysis of IR excess disc emission by \cite*{Schneider12a} suggests that none of these stars have appreciable excess emission and therefore this is unlikely to be the cause of the apparent luminosity spread.

An alternative possibility is that the apparent luminosity spread could be a consequence of a real age spread within the association. \cite{Weinberger13} propose that an age spread is probable given that the width of the age distribution for TWA members exceeds that which would be expected for a population of stars formed in a single burst. To place stricter constraints on whether the data are consistent with a real age spread or not, a more comprehensive study, such as that of \cite{Reggiani11} in the Orion Nebula Cluster which accounted for uncertainties in the distance, spectral type, unresolved binarity, accretion and photometric variability, would have to be performed.

\subsection{32 Ori group}
\label{discussion_32_ori}

The 32 Ori group is fairly new and has not been as well-characterised
as the other groups in this study. As such, our isochronal age
represents the first definitive published age for the group.  The
group is located in northern Orion, centred near 32~Ori and Bellatrix
(although not containing the latter), with a distinctly high proper
motion ($\mu_{\alpha}$, $\mu_{\delta}$ $\simeq +7, -33\,\rm{mas\,yr^{-1}}$)
and radial velocity ($v_r \simeq -18.5\,\rm{km\,s^{-1}}$) compared to
the much more distant Ori~OB1 young stellar population.
It was first reported by \cite{Mamajek07}, and a \emph{Spitzer} IR
survey of the group reported by \cite{Shvonski10} found several
members to have dusty debris discs cooler than $\sim 200\,\rm{K}$, namely
HD~35499 (F4V),
HD~36338 (F4.5V),
HR~1807 (A0Vn),
TYC~112-917-1 (K4), and
TYC~112-1486-1 (K3).
\cite{Kharchenko13} recovered the cluster and estimated it to to
be at 95\,pc, with mean proper motion $\mu_{\alpha}$, $\mu_{\delta}$
$\simeq +10.0, -32.2\,\rm{mas\,yr^{-1}}$ ($\pm 0.8\,\rm{mas\,yr^{-1}}$), a
core radius of $\sim1.6\,\rm{pc}$, and an isochronal age of $32\,\rm{Myr}$.

Our age for 32~Ori is statistically indistinguishable from that for
the BPMG (22 and 24\,Myr respectively) and the coevality of these two
groups is also evident from the position of the main-sequence turn-on.
In the 32 Ori group, HD~36338 (spectral type F4.5V) lies in a region
of the CMD which \cite{Mamajek14} termed the `false ZAMS' i.e. in a
region between the penultimate luminosity minima and the final
luminosity minima corresponding to the ZAMS.
All stars cooler than HD~36338 are over-luminous with respect to both
the `false ZAMS' and ZAMS, and therefore appear to be pre-MS
stars.
Note that the spectral type at which the main-sequence turn-on occurs
is slightly later in 32 Ori than the BPMG (in which it occurs
approximately one spectral type earlier) despite the slightly younger
isochronal age for the group, however the $\pm3\,\rm{Myr}$ uncertainty
on both isochronal ages and the paucity of stars with spectral types
between F4.5 and F9 in 32 Ori makes such a direct comparison of only
limited validity.
A comprehensive characterisation of both the stellar and circumstellar
disc content of this group is currently underway (Shvonski et al. in
preparation) which should help to better constrain this important
epoch for terrestrial planet formation.

\section{Conclusions}
\label{conclusions}

In this study we present a self-consistent, absolute age scale for
eight young ($\lesssim 200\,\rm{Myr}$), nearby ($\lesssim
100\,\rm{pc}$) moving groups in the solar neighbourhood. We use
previous literature assessments to compile a list of member and
high-probability candidate members for each of the young groups in our
sample. Creating four sets of semi-empirical pre-MS model isochrones
based on the observed colours of young stars in the Pleiades, in
conjunction with theoretical corrections for the dependence on
log$\,g$, we combine these with the $\tau^{2}$ maximum-likelihood
fitting statistic to derive ages for each group using the $M_{V}, V-J$
CMD. Our final adopted ages for each group in our sample are:
$149^{+51}_{-19}\,\rm{Myr}$ for the AB Dor moving group,
$24\pm3\,\rm{Myr}$ for the $\beta$ Pic moving group (BPMG),
$45^{+11}_{-7}\,\rm{Myr}$ for the Carina association,
$42^{+6}_{-4}\,\rm{Myr}$ for the Columba association,
$11\pm3\,\rm{Myr}$ for the $\eta$ Cha cluster, $45\pm4\,\rm{Myr}$ for
the Tucana-Horologium moving group (Tuc-Hor), $10\pm3\,\rm{Myr}$ for
the TW Hya association, and $22^{+4}_{-3}\,\rm{Myr}$ for the 32 Ori
group. At this stage, we are uncomfortable assigning a final,
unambiguous age to Argus as it appears as though the stars in our
membership list for the association suffer from a high level of
contamination and hence may not represent a single, coeval
population. Comparing our isochronal ages to literature LDB ages,
which are currently only available for the BPMG and Tuc-Hor, we find
consistency between these two age diagnostics for both groups. This
consistency instills confidence that our self-consistent, absolute age
scale for young, nearby moving groups is robust and hence we suggest
that these ages be adopted for future studies of these groups.

\section*{Acknowledgements}

CPMB and EEM acknowledge support from the University of Rochester School of Arts and Sciences. EEM also acknowledges support from NSF grant AST-1313029. We thank Lison Malo for discussions regarding the Bayesian analysis of membership probabilities for the young moving groups in this study and Alex Binks for his tentative upper age limit for the AB Dor moving group. This research has made use of archival data products from the Two-Micron All-Sky Survey (2MASS), which is a joint project of the University of Massachusetts and the Infrared Processing and Analysis Center, funded by the National Aeronautics and Space Administration (NASA) and the National Science Foundation. This research has also made extensive use of the VizieR and SIMBAD services provided by CDS as well as the Washington Double Star Catalog maintained at the U.S. Naval Observatory and the Tool for OPerations on Catalogues And Tables (TOPCAT) software package \citep{Taylor05}.

\bibliographystyle{mn3e}
\bibliography{references}

\appendix

\section{Between a rock and a hard place: de-constructing photometry of higher-order multiple systems}
\label{photometry_split}

\subsection{Methodology}
\label{methodology_split}

To de-construct photometry of higher-order multiple systems into individual components measurements we adopt the technique of \cite{Mermilliod92} which, in our case, uses the system $V$-band magnitude, system $V-J$ colour and $\Delta V$-band magnitude between the two components.

We illustrate this technique using the TWA member TWA~4 as an example. TWA~4 is a quadruple system (unresolved spectral type of K6IVe; \citealp{Pecaut13}) comprising two visual components (separation $< 1\,\rm{arcsec}$), which themselves are binary systems: A is a single-lined spectroscopic binary and B is a double-lined spectroscopic binary. Of the four components (designated Aa, Ab, Ba and Bb), only light from three of these -- Aa, Ba and Bb -- is visible in the combined optical spectrum \citep{Soderblom98}. We adopt the system $V$-band magnitude of $V_{\rm{AB}}=8.912\pm0.053\,\rm{mag}$ based on 601 observations by the ASAS. The 2MASS PSC gives a system $J$-band magnitude of $J_{\rm{AB}}=6.397\pm0.020\,\rm{mag}$, yielding a system $V-J$ colour of $(V-J)_{\rm{AB}}=2.515\pm0.057\,\rm{mag}$. For the difference in $V$-band magnitude between the A and B components we adopt $\Delta V_{\rm{A,B}}=0.504\pm0.030\,\rm{mag}$ from \cite{Soderblom98} based on two separate $V$-band measurements.

Whilst the \cite{Mermilliod92} formalism is demonstrated for splitting $BV$ photometry, this can be generalised to any combination of photometric bandpasses so long as the relationship between the colour and the magnitude is known (in our case the trend between $M_{V}$ and $V-J$). We therefore fit an empirical (linear) slope to the resolved TWA members in the $M_{V}, V-J$ CMD, for which we calculate a slope $a=1.530$, and derive the colour and magnitude of the primary component according to

\begin{equation}
(V-J)_{\rm{A}} = (V-J)_{\rm{AB}} + 2.5\,\mathrm{log}\frac{1+10^{-0.4(\Delta m)(1+a)/a}}{1+10^{-0.4\Delta m}}
\end{equation}

\noindent and

\begin{equation}
V_{\rm{A}} = V_{\rm{AB}} + 2.5\,\mathrm{log}(1+10^{-0.4\Delta m})
\end{equation}

\noindent where $\Delta m$ is the difference in the $V$-band magnitude between the two components. The colour and magnitude of the secondary component is then simply

\begin{equation}
(V-J)_{\rm{B}} = (V-J)_{\rm{A}} + \Delta m/a
\end{equation}

\noindent and

\begin{equation}
V_{\rm{B}} = V_{\rm{A}} + \Delta m.
\end{equation}

\noindent Hence for TWA~4 we derive, for the two visual components A and B, colours and magnitudes of $(V-J)_{\rm{A}}=2.399\pm0.057\,\rm{mag}$, $V_{\rm{A}}=9.442\pm0.061\,\rm{mag}$, $(V-J)_{\rm{B}}=2.729\pm0.057\,\rm{mag}$, and $V_{\rm{B}}=9.946\pm0.061\,\rm{mag}$. For the uncertainties on the individual component $V-J$ colours we assume the same value as that for the system $V-J$ colour, whereas for the individual $V$-band measurements we add the uncertainties on the system $V$-band and $\Delta V$-band measurements in quadrature.

\subsection{Other cases}
\label{other_cases_split}

\subsubsection{HD~20121}
\label{hd20121}

HD~20121 is a member of Tuc-Hor and consists of a triple system (A0V+F7III+F5V according to the Washington Double Star Catalog; \citealp{Mason01}) in which the A+B components are separated by less than $1\,\rm{arcsec}$ and the C component lies a further $3\,\rm{arcsec}$ away. We adopt the combined system $V$-band magnitude of $V_{\rm{ABC}}=5.923 \pm 0.005\,\rm{mag}$ from \cite{Mermilliod06} and a system $J$-band magnitude of $J_{\rm{ABC}}=5.118 \pm 0.030\,\rm{mag}$ from the 2MASS PSC, thereby defining a system $V-J$ colour of $(V-J)_{\rm{ABC}}=0.805 \pm 0.030\,\rm{mag}$. Tycho-2 provides resolved photometry for the combined A+B component and C component, from which we derive (using the relation of \citealp{Mamajek06} to transform $V_{\rm{T}}$ photometry into Johnson $V$) a difference between these two visual components of $\Delta V_{\rm{AB,C}}=2.642 \pm 0.036\,\rm{mag}$. Given the early spectral types of the components, in conjunction with the age of Tuc-Hor, we expect that all three components are on the main-sequence. Therefore, to split the photometry we fit an empirical (linear) fit to the A/F/G-type main-sequence relation of \citet[slope $a=3.311$]{Pecaut13} to derive individual component colours and magnitudes of $(V-J)_{\rm{AB}}=0.758 \pm 0.030\,\rm{mag}$, $(V-J)_{\rm{C}}=1.556 \pm 0.030\,\rm{mag}$, $V_{\rm{AB}}=6.014 \pm 0.036\,\rm{mag}$, and $V_{\rm{C}}=8.656 \pm 0.036\,\rm{mag}$.

The reader will note that the spectral types we provide in Table~\ref{tab:member_stars} for the three components of this system differ from those given in the Washington Double Star Catalog. The reason for this is simply that the spectral types quoted in the Catalog are inconsistent with the derived colours and magnitudes of the individual components. \cite{Hog00b} and \cite{Fabricius02} provide resolved Tycho-2 $(BV)_{\rm{T}}$ photometry for the individual components of this system. We use the conversion of \cite{Mamajek06} and the trigonometric parallax measurement of \cite{vanLeeuwen07} to calculate individual absolute $M_{V}$ magnitudes and Johnson $B-V$ colours, and after comparing these to the main-sequence relation of \cite{Pecaut13}, find that both metrics are consistent with spectral types of F4V+F9V+G9V.

\subsubsection{GJ~3322}
\label{gj3322}

GJ~3322 is a member of the BPMG and is described by \cite{Riedel14} as a triple system (unresolved spectral type of M4IVe) in which the A+C components form an unresolved binary and the B component is $\simeq 1\,\rm{arcsec}$ away. Adopting the combined system $V$-band magnitude of $V_{\rm{ABC}}=11.504 \pm 0.030\,\rm{mag}$ from UCAC4 and $J$-band magnitude of $J_{\rm{ABC}}=7.212 \pm 0.023\,\rm{mag}$ from the 2MASS PSC, we derive a system $V-J$ colour of $(V-J)_{\rm{ABC}}=4.292 \pm 0.038\,\rm{mag}$. \cite{Riedel14} quote a difference in magnitude between the combined A+C component and the B component of $\Delta K_{\rm{AC,B}}=1.03\,\rm{mag}$, where the $K$-band used is that of the CIT system. To estimate what the corresponding difference between the two components in the $V$-band is, we first assume that $\Delta K_{\rm{CIT}} \simeq \Delta K_{\rm{s}}$ and then fit an empirical (linear) fit to the BPMG members in the $M_{V}, V-K_{\rm{s}}$ CMD (for which we find a slope of $a=1.665$) to calculate $\Delta V_{\rm{AC,B}}=2.579\,\rm{mag}$ (for which we assume an uncertainty of 0.05\,mag). To split the photometry we fit an empirical (linear) fit to the BPMG members (slope $a=1.929$) to derive individual component colours and magnitudes of $(V-J)_{\rm{AC}}=4.225 \pm 0.038\,\rm{mag}$, $(V-J)_{\rm{B}}=5.562 \pm 0.038\,\rm{mag}$, $V_{\rm{AC}}=11.601 \pm 0.058\,\rm{mag}$, and $V_{\rm{B}}=14.180 \pm 0.058\,\rm{mag}$.

The recent study of \cite{Riedel14} also includes deblended photometric measurements for the individual components, yielding $V_{\rm{A}}=V_{\rm{C}}=12.46\,\rm{mag}$, $V_{\rm{B}}=13.56\,\rm{mag}$, $(V-J)_{\rm{A}}=(V-J)_{\rm{C}}=4.16\,\rm{mag}$, and $(V-J)_{\rm{B}}=4.91\,\rm{mag}$. \citeauthor{Riedel14} assume that the A and C components are of equal-mass, which is clearly not true as \cite{Delfosse99} provides radial velocity amplitudes. On the other hand, our method assumes that we are disentangling unresolved binaries, and therefore if one of the components is itself an unresolved binary, then our method is insufficiently robust to deal with the fact that one of the sources is unnaturally brighter because it is a multiple. As we do not have any information regarding the photometric difference between the A and C components, we opt to simply adopt the aforementioned values of \cite{Riedel14} for the three individual components of the system.

\section{Boot-strapped colours and questionable memberships}
\label{notes_on_individual_stars}

\subsection{Boot-strapped colours}
\label{boot-strapped_colours}

In Section~\ref{data:photometry} we discussed cases in which we were forced to infer $V-J$ colours for member stars from either the $V-K_{\rm{s}}$ or $B-V$ colour. We used the $V-K_{\rm{s}}$ colour when only the associated 2MASS PSC $J$-band Qflg was not `A'. To calculate the uncertainty on the inferred $V-J$ colour, we adopted the 2MASS PSC $K_{\rm{s}}$-band uncertainty and added this in quadrature with the $V$-band uncertainty. When the associated Qflgs of all three 2MASS PSC bandpasses were not `A' we used the $B-V$ colour to estimate $V-J$. In such cases, given that the 2MASS PSC photometry is unreliable, to assign an uncertainty on the $V-J$ colour, we assumed an uncertainty of 0.03\,mag combined in quadrature with the $V$-band uncertainty.

The following stars have $V-J$ colours inferred from their $V-K_{\rm{s}}$ colours: 32~Ori, AK~Pic, HD~31647A, HIP~88726, HR~136, HR~789, HR~1190, HR~1189, HR~1474, HR~1621, HR~2466, HR~6070, HR~7329, HR~7736, HR~8352, HR~8911, HR~9016, and HR~9062.

The following stars have $V-J$ colours inferred from their $B-V$ colours: HD~31647B, HR~126, HR~127, HR~674, HR~806, HR~838, HR~1249, HR~2020, HR~4023, HR~4534, HR~7012, HR~7348, HR~7590, HR~7790, and HR~8425.

\subsection{Questionable membership}
\label{questionable_membership}

In this Section we discuss those stars for which membership to a given moving group is either questionable or unlikely.

\subsubsection{AB Dor moving group}
\label{ab_dor_membership}

HR~7214 was designated a member of the AB Dor moving group by \cite{Zuckerman11}, however \cite{Malo13} derives a membership probability (including radial velocity and parallax information) of $P_{v+\pi}=80$~per~cent for this object due to its seemingly anomalous space motion $U$ in comparison to the bulk motion of the group as defined by bonafide members. We retain HR~7214 for our age analysis on the basis that there is no categoric evidence that is should be discarded. 

\subsubsection{BPMG}
\label{beta_pic_membership}

HR~789 is a tight spectroscopic binary which also has a more distance ($25\,\rm{arcsec}$) co-moving M dwarf companion (2MASS J02394829-4253049). This system was suggested as a member of Columba by \cite{Zuckerman11}, however \cite{Malo13} find it is more likely a BPMG member ($P_{v+\pi}=96$~per~cent). We therefore include HR~789 and 2MASS J02394829-4253049 as BPMG members for our age analysis.

GJ~9303 (HIP~47133) was proposed as a BPMG member by \cite{Schlieder12a}, however the Bayesian analysis of \cite{Malo13} suggests that it is more likely associated with the field ($P_{v+\pi}=99$~per~cent). The main reason for this assignment is that its Galactic position in $Z$ is far ($\sim 40\,\rm{pc}$) from the centre of bonafide BPMG members. With no available Li diagnostic to infer its youth, we discard GJ~9303 from our age analysis.

HR~7329 was classified as a BPMG member by \cite{Zuckerman01}, however \cite{Malo13} demonstrates that the radial velocity is discrepant ($\Delta v \simeq 14\,\rm{km\,s^{-1}}$) with regards to that predicted for the BPMG, and that if it is included in the prior the probability $P_{v+\pi}$ becomes zero. \cite{Malo13} suggest that either the close-by low-mass companion may affect the systematic velocity or that the radial velocity is erroneous due to the fast rotation of the primary ($v\,\rm{sin}i=330\,\rm{km\,s^{-1}}$). With no observations currently available to categorically rule out its association with the BPMG, we retain HR~7329 in the age analysis.

\subsubsection{Carina}
\label{carina_membership}

AB~Pic (HD~44627) has an ambiguous membership having previously been assigned to Tuc-Hor by \cite{Zuckerman04a} only to be revised to Carina by \cite{Torres08}. The Bayesian analysis of \cite{Malo13} suggests that it is more likely a member of the latter ($P_{v+\pi}=71$~per~cent for Carina compared with only 29~per~cent for Columba and 0~per~cent for Tuc-Hor). The higher probability that it belongs to Carina is a result of its Galactic positions $YZ$ which are more akin to that of Carina than Tuc-Hor. Furthermore, the reason it has a higher probability of being a Columba member (as compared to Tuc-Hor) is because there is a difference of only $1\,\rm{km\,s^{-1}}$ between the predicted radial velocities for Columba and Carina \citep{Malo13}. Until additional radial velocity measurements are made to categorically demonstrate that it is not a member of Carina, we retain AB~Pic as a member for our age analysis.

\subsubsection{Columba}
\label{columba_membership}

HD~15115 was previously believed to be a member of the BPMG (see e.g. \citealp{Moor06}), however \cite{Malo13} suggests that it is more likely to be a member of Columba ($P_{v+\pi}=87$~per~cent) and therefore we retain this as a member of Columba for the age determination. Similarly, whilst HD~23524 was originally designated a Tuc-Hor member by \cite{Zuckerman11}, \cite{Malo13} derive a probability of it being a Columba member of $P_{v+\pi}=98$~per~cent, and thus we include it in our age determination of Columba. In addition, we also include V1358~Ori (BD-03~1386) and DK~Leo (GJ~2079; originally assigned memberships in Tuc-Hor and the BPMG respectively) in our age analysis of Columba, as \cite{Malo13} derive membership probabilities of $P_{v+\pi}=85$ and 93~per~cent respectively.

The membership status of AS~Col (HD~35114) is ambiguous primarily as a result of a poorly constrained radial velocity (see e.g. \citealp*{Bobylev06} and \citealp{Bobylev07} who derive values of $15.2\pm1.6\,\rm{km\,s^{-1}}$ and $23.9\pm2.2\,\rm{km\,s^{-1}}$ respectively). The analysis of \cite{Malo13} finds that AS~Col belongs to either Columba or Tuc-Hor ($P_{v+\pi}=99$ per cent in both cases) depending on which radial velocity is adopted. Using the Bayesian Analysis for Nearby Young AssociatioNs (BANYAN; see \citealp{Malo13}) for AS~Col, but removing the radial velocity measurement as a prior, we find that the star appears to be a bona fide member of Columba. We therefore retain AS~Col as a Columba member for our age analysis of the association.

$\kappa$~And (HR~8976) is designated a bona fide member of Columba by \cite{Malo13} with a probability of $P_{v+\pi}=95$~per~cent. Recent evidence, however, suggests that $\kappa$~And is in fact much older ($\gtrsim 100\,\rm{Myr}$ depending on assumed composition; see e.g. \citealp{Hinkley13,Brandt15}). Given this evidence we exclude $\kappa$~And from the age analysis of Columba.

\subsubsection{TWA}
\label{tw_hydrae_membership}

We discard three stars (TWA~14, TWA~18 and TWA~31) from the age analysis of the TWA. The mean distance estimate of the group is $\sim 57\,\rm{pc}$ with a $1\sigma$ scatter of roughly $11-12\,\rm{pc}$ \citep{Mamajek05}, however both TWA~14 and TWA~18 have distances (trigonometric and kinematic respectively) of $\gtrsim 100\,\rm{pc}$ and thus if they were members, would represent $3\sigma$ outliers. In addition, we do not include TWA~31 in our age analysis due to its designation as a non-member by \cite{Ducourant14}.

\subsubsection{Tuc-Hor}
\label{tuc_hor_membership}

GJ~3054 (HIP~3556), HD~12894, HD~200798, and BS~Ind (HD~202947) are all designated members of Tuc-Hor by \cite{Zuckerman04a}, however \cite{Malo13} demonstrated that if their radial velocities are included as priors, their respective probabilities of belonging to Tuc-Hor significantly decrease due to a difference of $\simeq 7-8\,\rm{km\,s^{-1}}$ between the measured and predicted radial velocities. Without additional, higher precision, radial velocity measurements for these stars, we retain all four in our age determination of Tuc-Hor.

HR~943 was classified as a Tuc-Hor member by \cite{Zuckerman11}, however \cite{Malo13} find that it is equally likely to be a member of Tuc-Hor or Columba ($P_{v+\pi}=50$ and 49~per~cent respectively). Until further measurements are able to discriminate between which of the moving groups it belongs to, we HR~943 as a member of Tuc-Hor for our age analysis.

HR~6351 and V857~Ara (HD~155915) are given as Tuc-Hor members by \cite{Zuckerman11}, and although exhibiting signs of youth (e.g. circumstellar material and Li absorption), the analysis of \cite{Malo13} suggests that both are field objects. Without categoric evidence that these stars are either field dwarfs or young interlopers belonging to another co-moving group, we retain both stars as Tuc-Hor members in our age analysis.

\subsubsection{Duplicate members}
\label{duplicate_members}

Given that we have collated memberships from numerous different literature sources, it is possible that stars which have been assigned membership to one moving group in a particular study, may be classified as a member of a different group by another study. We identified a total of 6 duplicate stars, 5 of which are classified as members of Tuc-Hor and Columba by \cite{Kraus14} and \cite{Malo14a} respectively, namely CD-44~753 (2MASS~J02303239-4342232), 2MASS~J03050976-3725058, 2MASS~J04240094-5512223, 2MASS~J04515303-4647309 and 2MASS~J05111098-4903597. In each case the star has been assigned membership solely on the basis of the measured radial velocity, and for 4 out of the 5 cases (excepting CD-44~753) the two independent radial velocity measurements of \cite{Kraus14} and \cite{Malo14a} agree to within the quoted uncertainties. Given the additional prior information included in the Bayesian analysis of \citeauthor{Malo13}, we retain these stars as members of Columba, but note that the inclusion/exclusion of these stars for either group has a negligible effect on the best-fit age.

TYC~5853-1318-1 (2MASS~J01071194-1935359) was also found to be a duplicate, first suggested as a BPMG member by \cite{Kiss11}, but more recently advocated as a Tuc-Hor member by \cite{Kraus14}. The Bayesian analysis of \cite{Malo13} is unable to categorically assign membership to only one of the moving groups, instead suggesting it could belong to any of the BPMG, Tuc-Hor or Columba. Interestingly, \cite{Kraus14} only assigns membership to Tuc-Hor on the basis of strong Li absorption i.e. it is young. Its measured radial velocity ($v=9.3\pm0.5\,\rm{km\,s^{-1}}$) is somewhat discrepant with respect to that of the bulk of the other Tuc-Hor members ($\Delta v=8.25\,\rm{km\,s^{-1}}$), and on the basis of this should be considered a non-member (as stated in \citealp{Kraus14}). The measured radial velocity of \citet[$v=11.5\pm1.4\,\rm{km\,s^{-1}}$]{Kiss11} is consistent with that of \cite{Kraus14} to within the uncertainties. Furthermore, both of these velocities are consistent with that predicted by \cite{Malo13} if it is in fact a member of the BPMG. Based on this we prefer to assign membership of TYC~5853-1318-1 to the BPMG and include it in our age analysis of the group.

\section{Updated $\tau^{2}$ model for dealing with non-member contamination}
\label{tau2_non_members}

\subsection{Fitting datasets with non-members}
\label{fitting_datasets_with_non-members}

\subsubsection{Background}
\label{non_mem_background}

Non-members whose positions in the CMD lie outside the area of the cluster sequence can have an overwhelming effect on the fitted parameters, distorting them far away from the true values, or make the fit fail entirely. The reason is that the total likelihood (the logarithm of which is proportional to $\tau^{2}$) is the product of the likelihoods that the individual data points originate from the cluster sequence. Thus one datapoint with a likelihood close to zero for an otherwise good fit will drag the model towards it, or even give a probability of zero for the whole fit.

Conceptually the most straightforward solution is to have a model of the non-member contamination in CMD space. Perhaps surprisingly it turns out that a very crude model is effective; in \cite{Bell13} we used a uniform distribution over the area delineated by the maximum and minimum colours and magnitudes in the dataset (see also \citealp{vanDyk09}). In the same paper we showed how the uniformly distributed non-member model was formally equivalent to a soft-clipping scheme, and in fact actually used the latter in the fitting. However, here we will use the conceptual framework of a uniform non-member distribution. In part because it is better suited to the problem in hand, but also because it allows us to calculate a goodness-of-fit parameter, a step which was missing from our earlier soft-clipping technique.

Finally, before embarking on the formalism of this method, we should remark that whilst we will talk about non-members, in fact we should really refer to stars which do not fit our cluster model. For example, equal-mass triples will be very slightly above our equal-mass binary sequences, and hence in a region of zero probability because our model does not include higher order multiples. Therefore, such objects will be treated by our method as non-members, when in fact it is the model which is at fault. In this sense our uniform distribution is in part a Jaynes' fire extinguisher, a hypothesis which remains in abeyance unless needed by data which have a low probability of originating from our main hypothesis (see Section~4.4.1 of \citealp{Jaynes03}).

\subsubsection{Formalism and implementation}
\label{formalism_and_implementation}

As in \cite{Naylor06} we define the function to be minimised as

\begin{equation}
\tau^{2} = -2 \sum_{i=1,\,N} \mathrm{ln}  \iint U_{i}(c-c_{i}, m-m_{i})\, \rho (c,m)\, \mathrm{d}c\, \mathrm{d}m
\label{eqn:c1}
\end{equation}

\noindent (see also the elegant proof in \citealp{Walmswell13}). The model of the expected density of stars in the CMD is given by $\rho(c,m)$ and $U_{i}$ represents the uncertainties for the data point $i$. In terms of Bayes' theorem, the integral is $P(M)\,P(D|M )$, and so if we have two competing models this should be replaced by $P(M_{\mathrm{n}})\,P(D|M_{\mathrm{n}}) + P(M_{\mathrm{c}})\,P(D|M_{\mathrm{c}})$ where the subscripts c and n refer to cluster members and non-members respectively. If the probability that any given star is a member is given by $F_{i}$, then $P(M_{\mathrm{n}}) = 1-F_{i}$ and $P(M_{\rm c}) = F_{i}$.
Furthermore, $P(D|M_{\mathrm{n}})$ is the integral of $U_{i}\rho_{\mathrm{n}}$ and $P(D|M_{\mathrm{c}})$ the integral of $U_{i}\rho_{\mathrm{c}}$, where $\rho_{\mathrm{c}}$ is our usual cluster model and $\rho_{\mathrm{n}}$ is the model of the non-members. 
If the area delineated by the maximum and minimum colours and magnitudes in the dataset is $A$, then where $\rho_{\mathrm{n}}$ is non-zero $\rho_{\mathrm{n}} = 1/A$ since it must integrate to one (see \citealp{Naylor09}). 
Hence

\begin{equation}
\tau^{2} = -2 \sum_{i=1,\,N} \mathrm{ln} \left[ (1-F_{i})\rho_{\mathrm{n}} \iint U_{i} \mathrm{d}c\, \mathrm{d}m + F_{i} \iint U_{i}\rho_{\mathrm{c}} \mathrm{d}c\, \mathrm{d}m \right],
\label{eqn:c2}
\end{equation}

\noindent and thus

\begin{equation}
\tau^{2} = -2 \sum_{i=1,\,N} \mathrm{ln} \left[ \frac{1-F_{i}}{A} + F_{i} \iint U_{i}\rho_{\mathrm{c}} \mathrm{d}c\, \mathrm{d}m \right],
\label{eqn:c3}
\end{equation}

\noindent where we have used the facts that where $\rho_{\mathrm{n}}$ is non-zero in the CMD it is constant, and that $U_{i}$ integrates to one. Using this formula directly (rather than adding a constant to $\rho_{\mathrm{c}}$) is very straightforward to implement, since one simply calculates the likelihood in the normal way, adjusts it for the probability of membership and adds a constant. This has the advantage over our soft-clipping procedure that individual stars can be given different membership probabilities.

A further advantage of this formalism is that we can calculate how the position of a star in the CMD modifies our estimate of how likely it is to be a member.
If we apply Bayes' theorem to the hypothesis $M_{\rm{c}}$, that a star is a member of the cluster, which we are testing against a dataset $D$ then

\begin{eqnarray}
\label{eqn:c4}
P(M_{\mathrm{c}}|D) & = & \frac{P(M_{\mathrm{c}})P(D|M_{\mathrm{c}})}{P(D)} \\
   & = & \frac{P(M_{\mathrm{c}})P(D|M_{\mathrm{c}})}{P(M_{\mathrm{c}})P(D|M_{\mathrm{c}})+P(M_{\mathrm{n}})P(D|M_{\mathrm{n}})}. \nonumber
\end{eqnarray}

\noindent Using the same expressions for the probabilities and likelihoods we used for Eqn.~\ref{eqn:c3} we obtain

\begin{equation}
P_i(M_{\mathrm{c}}|D)={ {F_i\iint U_{i}\rho_{\mathrm{c}} \mathrm{d}c\, \mathrm{d}m }\over{ \frac{1-F_{i}}{A} + F_{i} \iint U_{i}\rho_{\mathrm{c}} \mathrm{d}c\, \mathrm{d}m }},
\label{eqn:c5}
\end{equation}

\noindent where in the denominator we have used the same simplifications as between Eqns.~\ref{eqn:c2} and \ref{eqn:c3}.
The intuitive interpretation of this equation is that at any point in the CMD the ratio of the model densities for members and non-members gives the membership probability for a star at that position, were its uncertainties in colour and magnitude infinitely small.
Allowance for uncertainties is made by convolving the densities with the uncertainty function.

\subsection{Testing the prior membership probabilities}
\label{testing_prior_membership_probabilities}

If the sum of the prior and posterior membership probabilities are very different, this could be an indication that the priors were incorrect.
A second way of exploring this is described in Section \ref{fitting_the_cmds} where we examined the distribution of the individual values of $\tau^2$.
The cumulative plots in Fig.~\ref{fig:distrib} show how mis-classified members have values of $\tau^2$ far exceeding those predicted by the fitting process.
If we lower the maximum prior membership probability, then both the predicted and measured values of $\tau^2$ develop a pedestal at the value of $\tau^2$ corresponding to the maximum membership probability (again shown in Fig.~\ref{fig:distrib}).
The simplest way to achieve this is to multiply all the priors by a factor, which we call $P_{\rm max}$ as it then corresponds to the maximum membership probability any star can have.
We found that adjusting $P_{\rm max}$ until either the pedestals in Fig.~\ref{fig:distrib} matched, or the sums of the prior and posterior membership probabilities were roughly in agreement gave very similar answers for the best value of $P_{\rm max}$.
Given the number of fits we had to perform, we took the latter course as straightforward to implement in an automated procedure.

We emphasise that $P_{\rm max}$ should not be adjusted to obtain a reasonable value for the goodness-of-fit [$\Pr(\tau^2)$], as it is possible have a reasonable value of $\Pr(\tau^2)$ but still have systematic residuals.
This can be seen by examining the cumulative distribution plots for the individual values of $\tau^2$, which seem to be a good space in which to examine the quality of a fit, rather like the residual plots in more conventional fitting.
Hence there are two metrics for a good fit, the numerical one of the value of $\Pr(\tau^2)$, and the qualitative one of the match between the predicted and model distributions of $\tau^2$.
It is possible to adjust $P_{\rm max}$ to achieve a good value of $\Pr(\tau^2)$ , but if this does not achieve a change in shape of the $\tau^2$ distribution that brings data and prediction into agreement, the fit is probably incorrect.

\subsection{Calculation of $\Pr(\tau^{2})$}
\label{calculation_of_pr_tau2}

\cite{Naylor06} showed how to assess whether a given model was a good fit to the data [$\Pr(\tau^{2})$] by calculating the probability that a random dataset drawn from the model, when fitted to the model would have a value of $\tau^{2}$ which exceeded that for the observed dataset. The method presented for calculating this involved convolving the distributions of $\tau^{2}$ for each individual datapoint with the distributions for all other data points. This procedure has the advantage of being a numerical equivalent to the way the expression for $\Pr(\chi^{2})$ is derived, but for the data presented here there are two significant problems.
First the method was unacceptably slow for some of the data, this is because some of the uncertainties are very large leading to slow convolutions.
Second the method has to be modified to allow for non-members.

\subsubsection{The new method}
\label{new_tau2_method}

We therefore elected to use a more direct method to calculate $\Pr(\tau^{2})$ where we simply simulated a thousand observed datasets using the parameters of the best-fitting model, and calculated $\tau^2$ for each of them.
For the simulation of cluster members, the simulated stars should in principle be drawn randomly from the stellar mass function multiplied by any selection effects, such as magnitude limits.
In practice the selection effects are often poorly understood, and so the function from which the stars are drawn is best defined from the magnitude distribution of the real dataset. 
This problem is not unique to this method of calculating $\Pr(\tau^{2})$, an analogous problem exists for the method described in \cite{Naylor06}.

To calculate $\Pr(\tau^2)$ we began by simulating a cluster of a million stars with the parameters of the best-fitting model (practically one can use the CMD of $\rho$ created for the fitting process) and 
grouped the stars into a number of magnitude bins equal to the number of observed data points.
The boundaries between these bins were set at the mid-magnitudes between each observed datapoint.
We then simulated a data point from each magnitude bin by first using the prior probability of membership to assign it to either the cluster or the field.   
Non-members were simulated by placing the star randomly within the rectangular area defined by the maximum and minimum colours and magnitudes of the real dataset (and thus there is a small chance it may be placed within the cluster sequence).
Cluster members were simulated taking their colours and magnitudes from a simulated star randomly chosen from within the magnitude bin.
Since the bins are more closely spaced where there are many real data points, this means the distribution of data points in the simulated dataset broadly follows the magnitude distribution of real data points.
Each simulated star was then displaced from its model colour and magnitude using the uncertainties of the corresponding real data point.
Following this procedure for each magnitude bin resulted in a simulated observation, for which we could calculate $\tau^2$ in the normal way.
The distribution of 1000 values of $\tau^2$ calculated in this way could then be used to calculate $\Pr(\tau^{2})$ for the real observation.

Finally, we must allow for the effect of free parameters on the expected values of $\tau^2$, since they will be lower for the best-fitting model if there are more free parameters than in the case where a single model is fitted.
For the reasons discussed in \cite{Naylor09} we correct for this by scaling the values of $\tau^2$ associated with a particular $\Pr(\tau^2)$.
We first subtract the expectation value of $\tau^2$ from each of the values, multiply them by $\sqrt{2(N-n)}$, where $n$ here refers to the number of free parameters and $N$ to the number of data points, and then add back the the expectation value less $n$.
Note the scaling factor was incorrectly stated in \cite{Naylor09}, and was also incorrect in earlier versions of the code, though the corrections are so small that it does not significantly affect earlier results.

\subsubsection{The accuracy of $\Pr(\tau^2)$}
\label{accuracy_of_pr_tau2}

\begin{figure}
\centering
\includegraphics[width=\columnwidth]{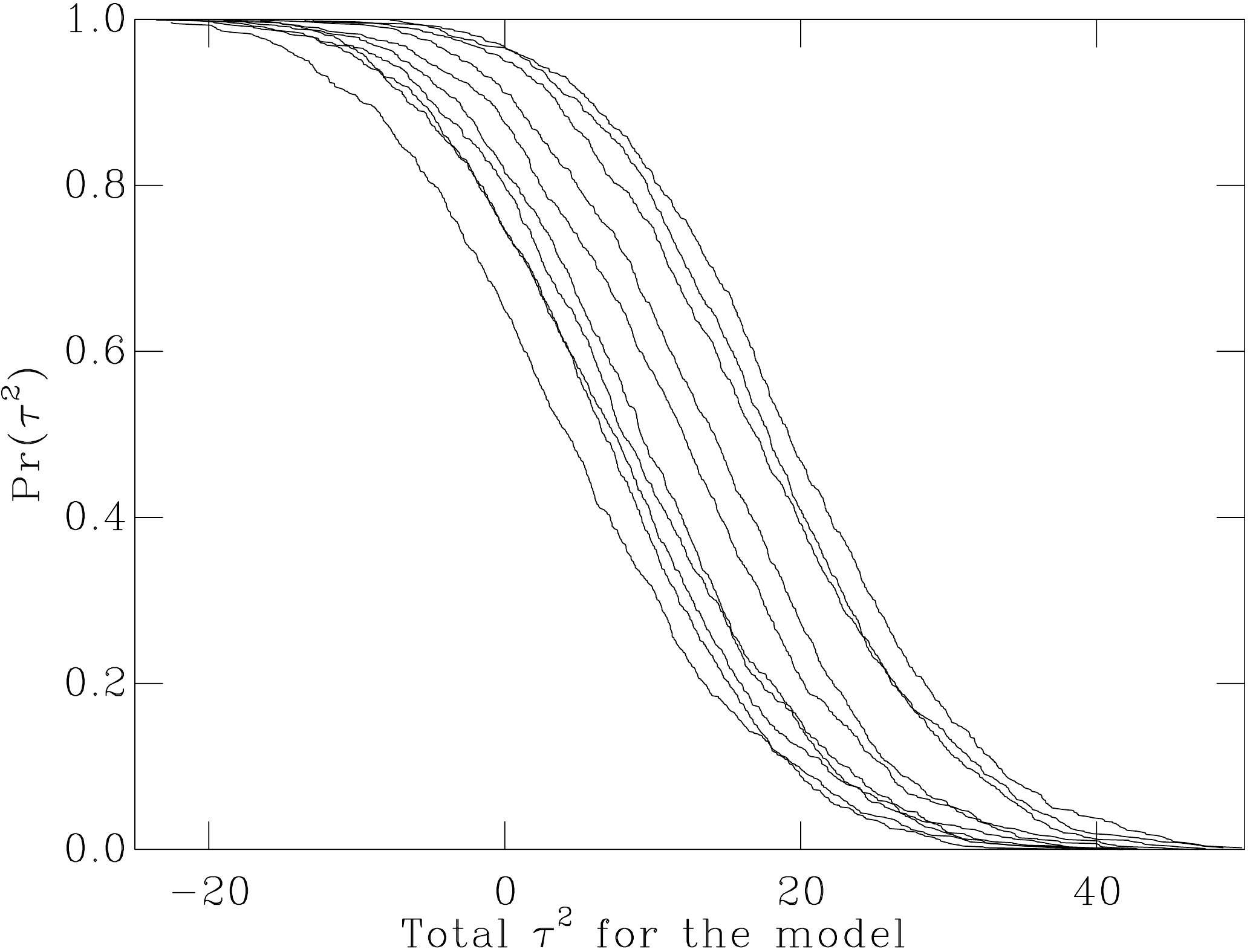}
\caption[]{The probability of a model exceeding a given value of $\tau^2$ calculated from ten different simulated observations of 30 stars from a model cluster of a million stars.}
\label{fig:tau2}
\end{figure}

In principle the cumulative distribution function for the expected values of $\Pr(\tau^2)$ is a function only of the best-fitting model and the way in which we select those stars which are observed.
If we have two different datasets consisting of different stars in the same cluster, provided those stars were selected using the same random process, $\Pr(\tau^2)$ would be the same for both datasets.
Unfortunately, the fact we do not know the selection criteria for the stars means that, as described above (see Appendix~\ref{new_tau2_method}) we have to rely on the observations themselves to give us the distribution in magnitudes expected for the sample, and so the value we calculate for $\Pr(\tau^2)$ depends on our sample.
This produces an uncertainly in our value of $\Pr(\tau^2)$, which it is important to quantify.
We performed an example simulation of 30 stars in a 10\,Myr-old cluster spread over 4 magnitudes in an $M_{V}, V-J$ CMD.
Fig.~\ref{fig:tau2} shows ten of the resulting distributions of $\tau^2$ which show that whilst the shape and the width of the distribution remain largely constant, the $\tau^2$ associated with a given probability varies with an RMS (calculated from 100 models) of $\simeq5$.
For good fits where $\Pr(\tau^2)\sim0.5$ this is not an issue since the 90 per cent width of the distribution is 30 in $\tau^2$, and so one would almost always  conclude the fit is a good one.
However, in the low-probability tail the test will show the fit is probably not a good one, but cannot give a useful answer as to how poor it is.

\label{lastpage}

\end{document}